\documentclass[preprint,prd,aps,showpacs,showkeys,nofootinbib]{revtex4}
\usepackage{graphicx}
\usepackage{dcolumn}
\usepackage{bm}
\usepackage{ulem}
\usepackage{color}
\usepackage[dvipsnames]{xcolor}
\usepackage{subfigure}
\usepackage{amssymb}
\usepackage{appendix}

\definecolor{light-gray}{gray}{0.78}
\definecolor{mid-gray}{gray}{0.55}
\definecolor{dark-gray}{gray}{0.32}
\begin{document}

\title{The lepton flavor universality including the $b\rightarrow c l \nu$ process in the $U(1)_X$SSM}
\author{Meng-Zi Cao$^{1,2,3}$, Shu-Min Zhao$^{1,2,3}$\footnote{zhaosm@hbu.edu.cn},Yue-Tong Liu$^{1,2,3}$ , Shuang Di$^{1,2,3}$ , Rong-Zhi Sun$^{1,2,3}$ , Xing-Xing Dong$^{1,2,3,4}$\footnote{dongxx@hbu.edu.cn}}

\affiliation{$^1$ Department of Physics, Hebei University, Baoding 071002, China}
\affiliation{$^2$ Hebei Key Laboratory of High-precision Computation and Application of Quantum Field Theory, Baoding, 071002, China}
\affiliation{$^3$ Hebei Research Center of the Basic Discipline for Computational Physics, Baoding, 071002, China}
\affiliation{$^4$ Departamento de Fisica and CFTP, Instituto Superior T$\acute{e}$cnico, Universidade de Lisboa,
Av.Rovisco Pais 1,1049-001 Lisboa, Portugal}
\date{\today}
\date{\today}

\begin{abstract}

  This paper calculates the lepton flavor universality including the $b\rightarrow c\ell \nu$ process under the U(1) extension of the minimal supersymmetric standard model ($U(1)_X$SSM), and provides numerical analysis
  and
  summary. The numerical results are used to create one-dimensional and multi-dimensional plots to analyze the impacts of various parameters on the  ratios $(\frac{R_{J/\psi}}{R^{SM}_{J/\psi}}$,
  $\frac{R_{{D}_s}}{R^{SM}_{D_s}},
\frac{R_{{D^*}_s}}{R^{SM}_{D^*_s}},\frac{R_{{\Lambda}_c}}{R^{SM}_{\Lambda_c}})$. The results show that several parameters have a certain effect on the ratios.
Although it can't match the experimental data very well, with the right combination of parameter values, it can give better numerical results than the Standard Model. This shows that the $U(1)_X$SSM  is very helpful for
studying
$b\rightarrow c\ell \nu$  processes.
\end{abstract}


\maketitle

\section{introduction}
The Standard Model (SM) is a very successful theoretical model for describing particles, but it still has some hard-to-solve problems.
Lepton Flavor Universality (LFU) stands as a cornerstone hypothesis of the SM, postulating that the electroweak gauge interactions are identical for all three generations of charged leptons($e,\mu,\tau$),
differing only by mass effects. Precision tests of LFU are therefore of importance, as any observed deviation would provide unambiguous evidence for physics beyond the SM (BSM). In recent years, the flavor
physics community has been captivated by persistent anomalies in $b$ -hadron decays, particularly in the charged-current transitions $ b\rightarrow{cl\nu}$.
The most prominent observables in this sector are the ratios$(\frac{R_{J/\psi}} {R^{SM}_{J/\psi}}$and$\frac{R_{{D^*}_s}}{R^{SM}_{D^*_s}}$)
, defined as the branching ratio of decays involving a  $\tau$ lepton normalized to those involving light leptons ($l=e,\mu$). While recent measurements from LHCb\cite{1}
 and Belle \cite{2,3,4} have reduced the tension with SM predictions, the experimental uncertainties remain significant.

 These branches are very important and are some key tools in the study of new physics(NP)\cite{5}. The current landscape suggests that if NP exists, it likely couples preferentially to the third generation of
 fermions due to the mass hierarchy\cite{6}.
To interpret these potential deviations model-independently, the Low-Energy Effective Field Theory (LEFT) provides a robust framework\cite{7}. In Ref\cite{6}, the effects of heavy BSM particles can be encapsulated in
Wilson coefficients (WCs) of dimension-six operators. Specifically, the $ b\rightarrow{cl\nu}$
transition can be mediated by scalar, vector, and tensor operators in addition to the SM $V-A$ current\cite{8}.

 The projected sensitivities of future colliders to these WCs are crucial for narrowing down the parameter space of specific UV-complete models\cite{9}.
Among various BSM scenarios, the $U(1)$ extension of the Supersymmetric Standard Model ($U(1)_X$SSM)\cite{10,11,12} is a compelling candidate\cite{11,13}. This model extends the MSSM by an additional $U(1)$ gauge symmetry,
introducing a heavy neutral gauge boson $Z^\prime$ and an extended Higgs sector\cite{11}. Unlike the MSSM, the $U(1)_X$SSM can naturally accommodate the observed Higgs mass and provide rich flavor
structures\cite{14}. In the context of $b\rightarrow{cl\nu}$ transitions,
$U(1)_X$SSM has more particles to produce new contributions.
In the $U(1)_X$SSM, we have research $R_D$ and $R_{D^*}$ in the previous work\cite{15}.
While the High-Luminosity LHC and Belle II will continue to lead the search for LFU violation, the upcoming Electron-Ion Collider (EIC) offers a unique and complementary environment\cite{16}.

Ref.\cite{6} shows that the future Electron-Ion
Collider(EIC) can produce abundant b-hadrons via deep inelastic scattering (DIS) at $\sqrt{s}\approx100\,${GeV}, and its clean polarized electron-proton collisions allow precise reconstruction of $\tau$-related decay
chains.
They discuss that EIC can constrain  LEFT Wilson coefficients to $\mathcal{O}(0.1)$ order through $R_{\Lambda_c}$ measurements\cite{17}, a capability unavailable to other facilities\cite{18}. There are also other works
relating to the
process $b\rightarrow{cl\nu}$ \cite{19,20,21,22,23,24}.
   Other U(1) extended supersymmetric constructions with right-handed neutrinos share partial structural similarities with our setup \cite{25,26}.
   Global correlations across all $b\rightarrow c\tau\nu$ exclusive channels
    in models possessing new physics have been systematically analyzed in prior literature \cite{27,28}.  In this work, we calculate the contributions to the  $b\rightarrow{cl\nu}$  in the $U(1)_X$SSM, and obtain the
    corresponding Wilson coefficients.
   We use the experiment constraints for these processes
   to constrain the $U(1)_X$SSM parameter space, specifically the couplings and masses of the extra gauge and Higgs bosons.

The paper is organized as follows.
We mainly introduce the content of $U(1)_X$SSM in section 2.
 In section 3, we deduce the formulas
 for the decays of $b\rightarrow c l\nu$ in the $U(1)_X$SSM. In
section 4, we give the numerical analysis, and the summary is given in section 5.
\section{the $U(1)_X$SSM}
This chapter will give an introduction to the $U(1)_X$SSM model. $U(1)_X$SSM  is a $U(1)$ gauge extension of the Minimal Supersymmetric Standard Model (MSSM). The core idea is to introduce a new $U(1)_X$ Abelian gauge
symmetry on base of the
MSSM, addressing issues like the $\mu$ problem and neutrino masses\cite{11}.  The $\mu$ problem in the MSSM is alleviated through the term $\lambda_{H}\hat{S}{\hat{H}_u}{\hat{H}_d}$, after the $S$ field obtains
  the vacuum expectation value $v_S$. MSSM lacks right-handed neutrinos, so neutrinos are massless. $U(1)_X$SSM introduces three generations of right-handed neutrino superfields $\hat{\nu}_i$, which generate tiny neutrino
  masses at tree level
via the seesaw mechanism\cite{23}, explaining neutrino oscillations.  The lightest  sneutrino or neutralino can serve as cold dark matter candidate\cite{21}.  The neutral CP-even parts of the new singlet Higgs superfields (
$\hat{\eta}$ , $\hat{\bar{\eta}}$ , $\hat{S}$ ) mix with MSSM Higgs ($H_u,H_d$) fields, boosting the Higgs mass at tree level to better match experimental values. The  gauge group of the $U(1)_X$SSM is $SU(3)_C\otimes.
SU(2)_L \otimes U(1)_Y\otimes U(1)_X$.   The quantum numbers of the particles in $U(1)_X$SSM are given in the table~\ref{quarks}\cite{29}.
\begin{table}
\caption{ The superfields in $U(1)_X$SSM}
\begin{tabular}{|c|c|c|c|c|}
\hline
Superfields & $SU(3)_C$ & $SU(2)_L$ & $U(1)_Y$ & $U(1)_X$ \\
\hline
$\hat{Q}_i$ & 3 & 2 & 1/6 & 0 \\
\hline
$\hat{u}^c_i$ & $\bar{3}$ & 1 & -2/3 & -$1/2$ \\
\hline
$\hat{d}^c_i$ & $\bar{3}$ & 1 & 1/3 & $1/2$  \\
\hline
$\hat{L}_i$ & 1 & 2 & -1/2 & 0  \\
\hline
$\hat{e}^c_i$ & 1 & 1 & 1 & $1/2$  \\
\hline
$\hat{\nu}_i$ & 1 & 1 & 0 & -$1/2$ \\
\hline
$\hat{H}_u$ & 1 & 2 & 1/2 & 1/2\\
\hline
$\hat{H}_d$ & 1 & 2 & -1/2 & -1/2 \\
\hline
$\hat{\eta}$ & 1 & 1 & 0 & -1 \\
\hline
$\hat{\bar{\eta}}$ & 1 & 1 & 0 & 1\\
\hline
$\hat{S}$ & 1 & 1 & 0 & 0 \\
\hline
\end{tabular}
\label{quarks}
\end{table}
The superpotential for this model reads:
\begin{eqnarray}
&&W=l_W\hat{S}+\mu\hat{H}_u\hat{H}_d+M_S\hat{S}\hat{S}-Y_d\hat{d}\hat{q}\hat{H}_d-Y_e\hat{e}\hat{l}\hat{H}_d+\lambda_H\hat{S}\hat{H}_u\hat{H}_d
\nonumber\\&&+\lambda_C\hat{S}\hat{\eta}\hat{\bar{\eta}}+\frac{\kappa}{3}\hat{S}\hat{S}\hat{S}+Y_u\hat{u}\hat{q}\hat{H}_u+Y_X\hat{\nu}\hat{\bar{\eta}}\hat{\nu}
+Y_\nu\hat{\nu}\hat{l}\hat{H}_u.
\end{eqnarray}
The soft SUSY breaking terms are
\begin{eqnarray}
&&\mathcal{L}_{soft}=\mathcal{L}_{soft}^{MSSM}-B_SS^2-L_SS-\frac{T_\kappa}{3}S^3-T_{\lambda_C}S\eta\bar{\eta}
+\epsilon_{ij}T_{\lambda_H}SH_d^iH_u^j\nonumber\\&&
-T_X^{IJ}\bar{\eta}\tilde{\nu}_R^{*I}\tilde{\nu}_R^{*J}
+\epsilon_{ij}T^{IJ}_{\nu}H_u^i\tilde{\nu}_R^{I*}\tilde{l}_j^J
-m_{\eta}^2|\eta|^2-m_{\bar{\eta}}^2|\bar{\eta}|^2\nonumber\\&&
-m_S^2S^2-(m_{\tilde{\nu}_R}^2)^{IJ}\tilde{\nu}_R^{I*}\tilde{\nu}_R^{J}
-\frac{1}{2}\Big(M_X\lambda^2_{\tilde{X}}+2M_{BB^\prime}\lambda_{\tilde{B}}\lambda_{\tilde{X}}\Big)+h.c~~.
\end{eqnarray}
Here, $v_u,~v_d,~v_\eta$,~ $v_{\bar\eta}$ and $v_S$  are the vacuum expectation values(VEVs) corresponding to the Higgs superfields $H_u$, $H_d$, $\eta$, $\bar{\eta}$ and $S$.
The three Higgs singlets are represented as
\begin{eqnarray}
&&\eta={1\over\sqrt{2}}\Big(v_{\eta}+\phi_{\eta}^0+iP_{\eta}^0\Big),~~~~~~~~~~~~~~~
\bar{\eta}={1\over\sqrt{2}}\Big(v_{\bar{\eta}}+\phi_{\bar{\eta}}^0+iP_{\bar{\eta}}^0\Big),\nonumber\\&&
\hspace{4.0cm}S={1\over\sqrt{2}}\Big(v_{S}+\phi_{S}^0+iP_{S}^0\Big).
\end{eqnarray}
The two Higgs doublets are represented as
\begin{eqnarray}
&&H_{u}=\left(\begin{array}{c}H_{u}^+\\{1\over\sqrt{2}}\Big(v_{u}+H_{u}^0+iP_{u}^0\Big)\end{array}\right),
~~~~~~
H_{d}=\left(\begin{array}{c}{1\over\sqrt{2}}\Big(v_{d}+H_{d}^0+iP_{d}^0\Big)\\H_{d}^-\end{array}\right),
\nonumber\\
\end{eqnarray}
The two angles used in the above  are defined as $\tan\beta=v_u/v_d$  and $\tan\beta_\eta=v_{\bar{\eta}}/v_{\eta}$.
 The definition of
$\tilde{\nu}_L$ and $\tilde{\nu}_R$ is
\begin{eqnarray}
\tilde{\nu}_L=\frac{1}{\sqrt{2}}\phi_l+\frac{i}{\sqrt{2}}\sigma_l,~~~~~~~~~~\tilde{\nu}_R=\frac{1}{\sqrt{2}}\phi_R+\frac{i}{\sqrt{2}}\sigma_R.
\end{eqnarray}
A new effect called canonical dynamical mixing is caused by the combination of two Abelian groups $U(1)_Y$ and $U(1)_X$.  Even if it's set to zero under the grand unified theory, this effect can still be induced through the
renormalization group equations (RGEs).
What's special is that the two Abelian gauge groups are indivisible, so we have a new way to use rotation matrices $R$($R^T$$R=1$)\cite{30,31} to change the basis:
\begin{eqnarray}
&&D_\mu=\partial_\mu-i\left(\begin{array}{cc}Y,&X\end{array}\right)
\left(\begin{array}{cc}g_{Y},&g{'}_{{YX}}\\g{'}_{{XY}},&g{'}_{{X}}\end{array}\right)
\left(\begin{array}{c}A_{\mu}^{\prime Y} \\ A_{\mu}^{\prime X}\end{array}\right)\;.
\label{gauge1}
\end{eqnarray}
 Among them, $A_{\mu}^{\prime Y}$ and $A^{\prime X}_\mu$ denote the gauge fields of $U(1)_Y$ and $U(1)_X$. We need to redefine it
 \begin{eqnarray}
&&R\left(\begin{array}{c}A_{\mu}^{\prime Y} \\ A_{\mu}^{\prime X}\end{array}\right)
=\left(\begin{array}{c}A_{\mu}^{Y} \\ A_{\mu}^{X}\end{array}\right)\;.
\label{gauge4}
\end{eqnarray}
 \begin{eqnarray}
&&\left(\begin{array}{cc}g_{Y},&g{'}_{{YX}}\\g{'}_{{XY}},&g{'}_{{X}}\end{array}\right)
R^T=\left(\begin{array}{cc}g_{1},&g_{{YX}}\\0,&g_{{X}}\end{array}\right)\;.
\label{gauge3}
\end{eqnarray}
Therefore, the covariant derivative\cite{30,31,32,33} of $U(1)_X$SSM becomes:
\begin{eqnarray}
&&D_\mu=\partial_\mu-i\left(\begin{array}{cc}Y,&X\end{array}\right)
\left(\begin{array}{cc}g_{1},&g_{{YX}}\\0,&g_{{X}}\end{array}\right)
\left(\begin{array}{c}A_{\mu}^{Y} \\ A_{\mu}^{X}\end{array}\right)\;.
\label{gauge1}
\end{eqnarray}


At the tree diagram level,
there are three neutral gauge bosons $A^{X}_\mu,~A^Y_\mu$ and $V^3_\mu$ mix together, and the mass matrix
is shown in the basis $(A^Y_\mu, V^3_\mu, A^{X}_\mu)$

\begin{eqnarray}
\left(\begin{array}{ccc}\frac{1}{8}g_{1}^{2}v^{2},
&-\frac{1}{8}g_{1}g_{2}v^{2},&\frac{1}{8}g_{1}(g_{YX}+g_{X})v^{2}\\
-\frac{1}{8}g_{1}g_{2}v^{2},
&\frac{1}{8}g_{2}^2v^{2},&-\frac{1}{8}g_{2}(g_{X}+g_{YX})v^{2}\\
\frac{1}{8}g_{1}(g_{YX}+g_{X})v^{2},&-\frac{1}{8}g_{2}
(g_{YX}+g_{X})v^{2},&\frac{1}{8}g_{X}^{2}\xi^{2}
+\frac{1}{8}(g_{YX}+g_{X})^2v^{2}\end{array}\right).
\end{eqnarray}

We use two mixing angles $\theta_{W}$ and $\theta_{W}'$ to diagonalize this matrix. $\theta_{W}$ is the Weinberg angle. While, $\theta_{W}'$ can be expressed as a new hybrid angle:
\begin{eqnarray}
\sin^2\theta_{W}'=\frac{1}{2}-\frac{(g_{{YX}}^2-g_{1}^2-g_{2}^2)v^2+
4g_{X}^2\xi^2}{2\sqrt{[(g_{{YX}}+g_{X})^2+g_{1}^2+g_{2}^2]^2v^4+8g_{X}^2[(g_{{YX}}+g_{X})^2-g_{1}^2-g_{2}^2]v^2\xi^2+16g_{X}^4\xi^4}}.
\end{eqnarray}

In addition, $\theta_{W}'$ will also exist in the coupling between $Z$ and $Z^{\prime}$. By deriving the mixed result of $A^{X}_\mu,~A^Y_\mu$ and $V^3_\mu$, we can get:
\begin{eqnarray}
&&\qquad\;\quad\;m_\gamma^2=0,\nonumber\\
&&\qquad\;\quad\;m_{Z,{Z^{'}}}^2=\frac{1}{8}\Big((g_{1}^2+g_2^2+g_{YX}^2)v^2+4g_{X}^2\xi^2 \nonumber\\
&&\qquad\;\qquad\;\qquad\;\mp\sqrt{(g_{1}^2+g_{2}^2+g_{YX}^2)^2v^4+8(g_{YX}^2-g_{1}^2-
g_{2}^2)g_{X}^2v^2\xi^2+16g_{X}^4\xi^4}\Big).
\end{eqnarray}

Based on $(\nu_L,\bar{\nu}_R)$, we derive the mass matrix of the neutrino
\begin{eqnarray}
M_{\nu}=
\left({\begin{array}{*{20}{c}}
0 & \frac{\upsilon_u}{\sqrt{2}}(Y_\nu^T)^{IJ}  \\
\frac{\upsilon_u}{\sqrt{2}}(Y_\nu)^{IJ} & \sqrt{2}\upsilon_{\bar{\eta}}(Y_X)^{IJ}  \\
\end{array}}
\right),
\end{eqnarray}
and it is diagonalized by the matrix $Z_\nu$ through the formula
\begin{eqnarray}
Z_\nu M_\nu Z^T_\nu=diag(M_\nu).
\end{eqnarray}
Based on $({\phi}_{l}, {\phi}_{r})$, we can write the mass-squared matrix of CP-even sneutrino as
\begin{eqnarray}
M^2_{\tilde{\nu}^R} = \left(
\begin{array}{cc}
m_{{\phi}_{l}{\phi}_{l}} &m^T_{{\phi}_{r}{\phi}_{l}}\\
m_{{\phi}_{l}{\phi}_{r}} &m_{{\phi}_{r}{\phi}_{r}}\end{array}
\right),\label{Rsneu}
 \end{eqnarray}
\begin{eqnarray}
&&m_{{\phi}_{l}{\phi}_{l}}= \frac{1}{8} \Big((g_{1}^{2} + g_{Y X}^{2} + g_{2}^{2}+ g_{Y X} g_{X})( v_{d}^{2}- v_{u}^{2})
+  g_{Y X} g_{X}(2 v_{\eta}^{2}-2 v_{\bar{\eta}}^{2})\Big)
\nonumber\\&&\hspace{1.8cm}+\frac{1}{2} v_{u}^{2}{Y_{\nu}^{T}  Y_\nu}  + m_{\tilde{L}}^2,
 \\&&m_{{\phi}_{l}{\phi}_{r}} = \frac{1}{\sqrt{2} } v_uT_\nu  +  v_u v_{\bar{\eta}} {Y_X  Y_\nu}
  - \frac{1}{2}v_d ({\lambda}_{H}v_S  + \sqrt{2} \mu )Y_\nu,\\&&
m_{{\phi}_{r}{\phi}_{r}}= \frac{1}{8} \Big((g_{Y X} g_{X}+g_{X}^{2})(v_{d}^{2}- v_{u}^{2})
+2g_{X}^{2}(v_{\eta}^{2}- v_{\bar{\eta}}^{2})\Big) + v_{\eta} v_S Y_X {\lambda}_{C}\nonumber \\&&\hspace{1.8cm}
 +m_{\tilde{\nu}}^2 + \frac{1}{2} v_{u}^{2}|Y_\nu|^2+  v_{\bar{\eta}} (2 v_{\bar{\eta}}Y_X  Y_X  + \sqrt{2} T_X).
\end{eqnarray}

The matrix is diagonalized through the transformation of the $Z^R$ matrix.
The mass squared matrix for CP-odd sneutrino $({\sigma}_{l}, {\sigma}_{r})$ is also deduced here:

\begin{eqnarray}
M^2_{\tilde{\nu}^I} = \left(
\begin{array}{cc}
m_{{\sigma}_{l}{\sigma}_{l}} &m^T_{{\sigma}_{r}{\sigma}_{l}}\\
m_{{\sigma}_{l}{\sigma}_{r}} &m_{{\sigma}_{r}{\sigma}_{r}}\end{array}
\right),
 \end{eqnarray}
\begin{eqnarray}
&&m_{{\sigma}_{l}{\sigma}_{l}}= \frac{1}{8} \Big((g_{1}^{2} + g_{Y X}^{2} + g_{2}^{2}+  g_{Y X} g_{X})( v_{d}^{2}- v_{u}^{2})
+  2g_{Y X} g_{X}(v_{\eta}^{2}-v_{\bar{\eta}}^{2})\Big)
\nonumber\\&&\hspace{1.8cm}+\frac{1}{2} v_{u}^{2}{Y_{\nu}^{T}  Y_\nu}  + m_{\tilde{L}}^2,
 \\&&m_{{\sigma}_{l}{\sigma}_{r}} = \frac{1}{\sqrt{2} } v_uT_\nu -  v_u v_{\bar{\eta}} {Y_X  Y_\nu}
  - \frac{1}{2}v_d ({\lambda}_{H}v_S  + \sqrt{2} \mu )Y_\nu,\\&&
m_{{\sigma}_{r}{\sigma}_{r}}= \frac{1}{8} \Big((g_{Y X} g_{X}+g_{X}^{2})(v_{d}^{2}- v_{u}^{2})
+2g_{X}^{2}(v_{\eta}^{2}- v_{\bar{\eta}}^{2})\Big)- v_{\eta} v_S Y_X {\lambda}_{C}\nonumber \\&&\hspace{1.8cm}
+m_{\tilde{\nu}}^2 + \frac{1}{2} v_{u}^{2}|Y_\nu|^2+  v_{\bar{\eta}} (2 v_{\bar{\eta}}Y_X  Y_X  - \sqrt{2} T_X).
\end{eqnarray}
The mass-squared matrix of the scalar lepton based on ($\tilde{e}_{L}$,$\tilde{e}_{R}$) is shown below
\begin{eqnarray}
m^2_{\tilde{e}} = \left(
\begin{array}{cc}
m_{{\tilde{e}_{L}}{\tilde{e}_{L}}^*}&\frac{1}{2}(\sqrt{2}v_{d}{T_e}^\dag-v_u({\lambda}_{H}v_S  + \sqrt{2} \mu ){Y_e}^\dag\\
\frac{1}{2}(\sqrt{2}v_{d}{T_e}-v_u{Y_e}(v_S{\lambda}_{H^*}  + \sqrt{2} \mu^* )&m_{{\tilde{e}_{R}}{\tilde{e}_{R}}^*}\end{array}
\right),
 \end{eqnarray}
\begin{eqnarray}
&&m_{{\tilde{e}_{L}}{\tilde{e}_{L}}^*}= m^2_{\tilde{L}} +\frac{1}{8} \Big((g_{1}^{2} + g_{Y X}^{2} - g_{2}^{2}+  g_{Y X} g_{X})( v_{d}^{2}- v_{u}^{2})
+  2g_{Y X} g_{X}(v_{\eta}^{2}-v_{\bar{\eta}}^{2})\Big)
\nonumber\\&&\hspace{1.8cm}+\frac{1}{2} v_{d}^{2}{Y_{e}^\dag Y_e},
 \\&&m_{{\tilde{e}_{R}}{\tilde{e}_{R}}^*} =m^2_{\tilde{E}} -\frac{1}{8} \Big[2(g_{1}^{2} + g_{Y X}) + 3 g_{Y X} g_{X}+ g_{X}^2]( v_{d}^{2}- v_{u}^{2})
+  4g_{Y X} g_{X}(v_{\eta}^{2}-v_{\bar{\eta}}^{2})\Big)
\nonumber\\&&\hspace{1.8cm}+\frac{1}{2} v_{d}^{2}Y_e{Y_{e}^\dag }
\end{eqnarray}
We can use the transpose matrix $Z^E$ to diagonalize the scalar lepton mass-squared matrix.\\
In the same way, we can also derive the mass matrices of other particles.\\
The mass matrix for neutralino in the basis $(\lambda_{\tilde{B}}, \tilde{W}^0, \tilde{H}_d^0, \tilde{H}_u^0,
\lambda_{\tilde{X}}, \tilde{\eta}, \tilde{\bar{\eta}}, \tilde{s}) $ is,
\begin{equation}
m_{\tilde{\chi}^0} = \left(
\begin{array}{cccccccc}
M_1 &0 &-\frac{g_1}{2}v_d &\frac{g_1}{2}v_u &{M}_{B B'} &0  &0  &0\\
0 &M_2 &\frac{1}{2} g_2 v_d  &-\frac{1}{2} g_2 v_u  &0 &0 &0 &0\\
-\frac{g_1}{2}v_d &\frac{1}{2} g_2 v_d  &0
&m_{\tilde{H}_u^0\tilde{H}_d^0} &m_{\lambda_{\tilde{X}}\tilde{H}_d^0} &0 &0 & - \frac{{\lambda}_{H} v_u}{\sqrt{2}}\\
\frac{g_1}{2}v_u &-\frac{1}{2} g_2 v_u  &m_{\tilde{H}_d^0\tilde{H}_u^0} &0 &m_{\lambda_{\tilde{X}}\tilde{H}_u^0} &0 &0 &- \frac{{\lambda}_{H} v_d}{\sqrt{2}}\\
{M}_{B B'} &0 &m_{\tilde{H}_d^0\lambda_{\tilde{X}}} &m_{\tilde{H}_u^0\lambda_{\tilde{X}}} &{M}_{BL} &- g_{X} v_{\eta}  &g_{X} v_{\bar{\eta}}  &0\\
0  &0 &0 &0 &- g_{X} v_{\eta}  &0 &\frac{1}{\sqrt{2}} {\lambda}_{C} v_S  &\frac{1}{\sqrt{2}} {\lambda}_{C} v_{\bar{\eta}} \\
0  &0 &0 &0 &g_{X} v_{\bar{\eta}}  &\frac{1}{\sqrt{2}} {\lambda}_{C} v_S  &0 &\frac{1}{\sqrt{2}} {\lambda}_{C} v_{\eta} \\
0 &0 & - \frac{{\lambda}_{H} v_u}{\sqrt{2}} &- \frac{{\lambda}_{H} v_d}{\sqrt{2}} &0 &\frac{1}{\sqrt{2}} {\lambda}_{C} v_{\bar{\eta}}
 &\frac{1}{\sqrt{2}} {\lambda}_{C} v_{\eta}  &m_{\tilde{s}\tilde{s}}\end{array}
\right),\label{neutralino}
 \end{equation}
\begin{eqnarray}
&& m_{\tilde{H}_d^0\tilde{H}_u^0} = - \frac{1}{\sqrt{2}} {\lambda}_{H} v_S  - \mu ,~~~~~~~
m_{\tilde{H}_d^0\lambda_{\tilde{X}}} = -\frac{1}{2} \Big(g_{Y X} + g_{X}\Big)v_d, \nonumber\\&&
m_{\tilde{H}_u^0\lambda_{\tilde{X}}} = \frac{1}{2} \Big(g_{Y X} + g_{X}\Big)v_u
 ,~~~~~~~~~~~~
m_{\tilde{s}\tilde{s}} = 2 M_S  + \sqrt{2} \kappa v_S.\label{neutralino1}
\end{eqnarray}
This matrix is diagonalized by $Z^N$
\begin{equation}
Z^{N^*} m_{\tilde{\chi}^0} Z^{N{\dagger}} = m^{diag}_{\tilde{\chi}^0}.
\end{equation}

Based on ($\tilde{W}^-$,$\tilde{H}_d^-$) and ($\tilde{W}^+$,$\tilde{H}_u^+$), the mass matrix of the chargino($\chi^\pm$) is
\begin{eqnarray}
m_{\chi^\pm} = \left(
\begin{array}{cc}
M_{2}&\frac{1}{\sqrt{2}} g_{2} v_u\\
\frac{1}{\sqrt{2}} g_{2} v_d&\frac{1}{\sqrt{2}}{\lambda}_{H} v_S +\mu\end{array}
\right).
 \end{eqnarray}
The above matrix is diagonalized by two unitary matrices $U$ and $V$.
\begin{eqnarray}
U^* m_{\chi^\pm} {V^\dag}=m^{diag}_{\chi^\pm}
\end{eqnarray}
This is the mass matrix of the scalar down quark ($\tilde{D}$) based on ($\tilde{d}_{{L},\alpha_{1}}$,$\tilde{d}_{{R},\alpha_{2}}$) and ($\tilde{d}^*_{{L},\beta_{1}}$,$\tilde{d}^*_{{R},\beta_{2}}$).
\begin{eqnarray}
M^2_{\tilde{D}} = \left(
\begin{array}{cc}
m_{{\tilde{d}_{L}}{\tilde{d}^*_{L}}} &m\dag_{{\tilde{d}_{R}}{\tilde{d}^*_{L}}}\\
m_{{\tilde{d}_{R}}{\tilde{d}^*_{L}}} &m_{{\tilde{d}_{R}}{\tilde{d}^*_{R}}}\end{array}
\right),\label{Rsneu}
 \end{eqnarray}
among them
\begin{eqnarray}
&&m_{{\tilde{d}_{L}}{\tilde{d}^*_{L}}} = \frac{1}{24} \Big((3g_{2}^{2} + g_{Y X}^{2} + g_{1}^{2}+  g_{Y X} g_{X})( v_{u}^{2}- v_{d}^{2})
+  2g_{Y X} g_{X}(v_{\bar{\eta}}^{2}-v_{\eta}^{2})\Big)
\nonumber\\&&\hspace{1.8cm}+m^2_{\tilde{Q}}+ \frac{v_{d}^{2}}{2} {Y_d}^2,
 \\&&m_{{\tilde{d}_{R}}{\tilde{d}^{*}_{L}}} = -\frac{1}{2} \Big(-\sqrt{2}({v_d} {T_d} +{v_u} {Y_d}\mu)+ {v_u}{v_S}{Y_d}{\lambda_H}\Big),\\&&
m_{{\tilde{d}_{R}}{\tilde{d}^*_{R}}}= \frac{1}{24} \Big((2g_{1}^{2} + 2g_{Y X}^{2} + 3g_{X}^{2}+  5g_{Y X} g_{X})( v_{u}^{2}- v_{d}^{2})
+  2(2g_{Y X} g_{X}\nonumber\\&&\hspace{1.8cm}+3g_{X}^{2})(v_{\bar{\eta}}^{2}-v_{\eta}^{2})\Big)
+m^2_{\tilde{D}}+ \frac{v_{d}^{2}}{2} {Y_d}^2.
\end{eqnarray}
This is the mass matrix of the scalar up quark ($\tilde{U}$) based on ($\tilde{u}_{{L},\alpha_{1}}^0$,$\tilde{u}_{{R},\alpha_{2}}$) and ($\tilde{u}_{{L},\beta_{1}}^*$,$\tilde{u}_{{R},\beta_{2}}^*$).
\begin{eqnarray}
M^2_{\tilde{U}} = \left(
\begin{array}{cc}
m_{{\tilde{u}_{L}}{\tilde{u}^*_{L}}} &m\dag_{{\tilde{u}_{R}}{\tilde{u}^*_{L}}}\\
m_{{\tilde{u}_{R}}{\tilde{u}^*_{L}}} &m_{{\tilde{u}_{R}}{\tilde{u}^*_{R}}}\end{array}
\right)
 \end{eqnarray}
among them
\begin{eqnarray}
&&m_{{\tilde{u}_{L}}{\tilde{u}^*_{L}}} = \frac{1}{24} \Big(( 4g_{1}^{2}-3g_{2}^{2} + g_{Y X}^{2} +  g_{Y X} g_{X})( v_{u}^{2}- v_{d}^{2})
+  2g_{Y X} g_{X}(v_{\bar{\eta}}^{2}-v_{\eta}^{2})\Big)
\nonumber\\&&\hspace{1.8cm}+m^2_{\tilde{Q}}+ \frac{v_{u}^{2}}{2} {Y_u}^2,
 \\&&m_{{\tilde{u}_{R}}{\tilde{u}^*_{L}}} = -\frac{1}{2} \Big(\sqrt{2}({v_d} {Y_u}\mu-{v_u} {T_u} +)+ {v_d}{v_S}{Y_u}{\lambda_H}\Big),\\&&
m_{{\tilde{u}_{R}}{\tilde{u}_{R}^*}}= \frac{1}{24} \Big((4g_{1}^{2} + 4g_{Y X}^{2} + 3g_{X}^{2}+  7g_{Y X} g_{X})( v_{d}^{2}- v_{u}^{2})
+  2(4g_{Y X} g_{X}+3g_{X}^{2}(v_{\eta}^{2}-v_{\bar{\eta}}^{2})\Big)
\nonumber\\&&\hspace{1.8cm}+m^2_{\tilde{U}}+ \frac{v_{u}^{2}}{2} {Y_u}^2.
\end{eqnarray}
The mass matrix of charged Higgs bosons ($\tilde{H^\pm}$) is based on (${H}_{d}^-$,${H}^{+,*}_{u}$) and (${H}^{-,*}_{d}$,${H}^+_{u}$).
\begin{eqnarray}
M^2_{H^-} = \left(
\begin{array}{cc}
m_{{H}_{d}^-{H}^{-,*}_{d}} &m^*_{{H}^{+,*}_{u}{H}^{-,*}_{d}}\\
m_{{H}^-_{d}{H}^+_{u}} &m_{{H}^{+,*}_{u}{H}^+_{u}}\end{array}
\right),\label{Rsneu}
 \end{eqnarray}
\begin{eqnarray}
&&m_{{H}_{d}^-{H}^{-,*}_{d}} = \frac{1}{8} \Big((g_{1}^{2}+g_{X}^{2})v_{d}^{2} + (g_{2}^{2}-g_{X}^{2})v_{u}^{2}+ (g_{Y X}^{2} +g_{1}^{2}) g_{Y X} )(v_{d}^{2} - v_{u}^{2})
\nonumber\\&&\hspace{1.8cm}+ 2g_{X}^{2}v_{\bar{\eta}}^{2}+2[ g_{Y X}g_{X}(v_{d}^{2}+ v_{\eta}^{2}- v_{\bar{\eta}}^{2}-v_{u}^{2})+2g_{X}^{2}v_{\eta}^{2}\Big)
\nonumber\\&&\hspace{1.8cm}+\Big(|\mu|^2+\sqrt{2}v_S\Re(\mu{\lambda}^*_{H})+\frac{1}{2}v_{S}^{2}|{\lambda}_{H}|^2\Big),
\\&&m_{{H}^-_{d}{H}^+_{u}} = \frac{1}{2} \Big(2({\lambda}_{H}{l_{W}}^*+B_{\mu})+{\lambda}_{H}(2\sqrt{2}
v_{S}M_{S}^*-v_{d}v_{u}{\lambda}_{H}^*+v_{\eta}v_{\bar{\eta}}{\lambda}_{C}^*\nonumber\\&&\hspace{1.8cm}+\sqrt{2}{v_{S}}{T_{{\lambda}_{H}}})\Big)
+\frac{1}{4}g_{2}^{2}v_{d}v_{u},
\\&&m_{{H}^{+,*}_{u}{H}^+_{u}}= \frac{1}{8}\Big((g_{2}^{2}-g_{X}^{2})v_{d}^{2}+(g_{2}^{2}+g_{X}^{2})v_{u}^{2}(g_{1}^{2}+g_{YX}^{2})(v_{u}^{2}-v_{d}^{2})
\nonumber\\&&\hspace{1.8cm}-2g_{X}^{2}v_{\eta}^{2}+2[g_{YX}g_{X}(v_{u}^{2}+v_{\bar{\eta}}^{2}-v_{d}^{2}-v_{\eta}^{2})+g_{X}^{2}v_{\bar{\eta}}^{2}]\Big)
\nonumber\\&&\hspace{1.8cm}+\frac{1}{2}\Big(2|\mu|^2+2\sqrt{2}v_S\Re(\mu{\lambda}^*_{H})+v_{S}^{2}|{\lambda}_{H}|^2).
\end{eqnarray}
Here are some coupling vertices needed in this article.
The coupling vertexes of lepton-chargino-CP-odd(CP-even) sneutrino are
\begin{eqnarray}
&&\mathcal{L}_{\tilde{\nu}^R\bar{l}\chi^-}=\bar{l}_{i}\Big\{\frac{1}{\sqrt{2}}U^*_{j2}Z^{R*}_{ki}Y_l^iP_L-\frac{1}{\sqrt{2}}g_2V_{j1}Z^{R*}_{ki}P_R\Big\}\chi_j^-\tilde{\nu}^R_k,
\\&&\mathcal{L}_{\tilde{\nu}^I\bar{l}\chi^-}=\bar{l}_{i}\Big\{\frac{i}{\sqrt{2}}U^*_{j2}Z^{I*}_{ki}Y_l^iP_L-\frac{i}{\sqrt{2}}g_2V_{j1}Z^{I*}_{ki}P_R\Big\}\chi_j^-\tilde{\nu}^I_k.
\end{eqnarray}
The coupling vertexes
of neutralino-neutrino-CP-odd(CP-even) sneutrino are
\begin{eqnarray}
&&\mathcal{L}_{\tilde{\nu}^R\nu\chi^0}=\frac{1}{2}\bar{\chi}_i^0\Big\{(-g_2N^{*}_{i2}+g_{YX}N^{*}_{i5}+g_1N^{*}_{i1})
\sum_{a=1}^3Z^{R*}_{ka}U_{ja}^{V*}P_L\nonumber\\&&\hspace{1.7cm}+
(-g_2N_{i2}+g_{YX}N_{i5}+g_1N_{i1})\sum_{a=1}^3Z^{R*}_{ka}U_{ja}^{V}P_R\Big\}\nu_j\tilde{\nu}^R_k.
\\&&\mathcal{L}_{\tilde{\nu}^I\nu\chi^0}=-\frac{i}{2}\bar{\chi}_i^0\Big\{(-g_2N^{*}_{i2}+g_{YX}N^{*}_{i5}+g_1N^{*}_{i1})
\sum_{a=1}^3Z^{I*}_{ka}U_{ja}^{V*}P_L\nonumber\\&&\hspace{1.7cm}+
(g_2N_{i2}-g_{YX}N_{i5}-g_1N_{i1})\sum_{a=1}^3Z^{I*}_{ka}U_{ja}^{V}P_R\Big\}\nu_j\tilde{\nu}^I_k.
\end{eqnarray}
The coupling vertex of neutrino-slepton-chargino is
\begin{eqnarray}
&&\mathcal{L}_{\nu\chi^\pm\tilde{L}}=\bar{\nu}_{i}\Big((-g_{2}U^*_{j1}\sum_{a=1}^3U_{ia}^{V*}Z^{E}_{ka}+U^{*}_{j2}\sum_{a=1}^3U_{ia}^{V*}Y^{a}_lZ^{E}_{k(3+a)})P_L
\nonumber\\&&\hspace{1.7cm}+
\sum_{a,b=1}^3Y^{ab}_{\nu}U_{i(3+a)}^{V}Z^{E}_{kb}V_{j2}P_R\Big)\chi_j^\pm\tilde{L}_k,
\end{eqnarray}
The coupling vertex of neutrino-slepton-lepton is
\begin{eqnarray}
&&\mathcal{L}_{\chi^0l\tilde{L}}=\bar{\chi}^0\Big\{\Big(\frac{1}{\sqrt{2}}(g_{1}N^*_{i1}+g_{2}N^*_{i2}+g_{YX}N^*_{i5})Z^{E}_{kj}-N^*_{i3}Y^{j}_{l}Z^{E}_{k(3+j)}\Big)P_L
\nonumber\\&&\hspace{1.7cm}-[\frac{1}{\sqrt{2}}\Big(2g_{1}N_{i1}+(2g_{YX}+g_X)N_{i5}\Big)Z^E_{k(3+j)}+Y^j_{l}Z^E_{kj}N_{i3}]P_R\Big\}l_{j}\tilde{L}_{k}.
\end{eqnarray}
Also, we use the vertices related to W boson as follows:
\begin{eqnarray}
&&\mathcal{L}_{\tilde{\nu}^{R*}\tilde{L}W}=-\frac{1}{2}g_2\tilde{L}_i\tilde{\nu}_{j}^{R*}\sum_{a=1}^3Z^{E*}_{ia}Z^{R*}_{ja}(-p_{u^j}^{\tilde{\nu}^R_j}+p^{\tilde{L}_i}_\mu)W^\mu,
\\&&\mathcal{L}_{\tilde{\nu}^{I*}\tilde{L}W}=\frac{i}{2}g_2\tilde{L}_i\tilde{\nu}_{j}^{I*}\sum_{a=1}^3Z^{E*}_{ia}Z^{I*}_{ja}(-p_{u^j}^{\tilde{\nu}^I_j}+p^{\tilde{L}_i}_\mu)W^\mu,
\\&&\mathcal{L}_{{{\chi}^0_{i}}{\chi^\pm_{j}}W}=
-\frac{1}{2}g_{2}{\bar\chi^0_j}
\Big[(2U^*_{j1}N_{i2}+\sqrt{2}U^*_{j2}N_{i3})
{\gamma_\mu}P_L+(2N^*_{i4}V_{j2}){\gamma_\mu}P_R\Big]{\chi^\pm_{j}}W^\mu.
\end{eqnarray}
The vertices related to quarks are as follows:
\begin{eqnarray}
&&\mathcal{L}_{{\chi}^0\tilde{D}d}=-\frac{i}{6}\bar{\chi}^0_{i}\Big\{\Big[\sqrt{2}(g_{1}N_{1i}
-3g_{2}N_{2i}+g_{YX}N_{5i})Z^{\tilde{D}^*}_{jk}
+6N_{3i}Y^{j}_{d}Z^{\tilde{D}^*}_{(3+j)k}\Big]P_L
\nonumber\\&&\hspace{1.6cm}+\Big[6Y^{j}_{d}Z^{\tilde{D}^*}_{jk}N^*_{3i}
+\sqrt{2}Z^{\tilde{D}^*}_{(3+j)k}[2g_{1}N^*_{1i}+(2g_{YX}+3g_{X})
N^*_{5i}]\Big]P_R\Big\}d_{j}\tilde{D}^*_{k},
\\&&\mathcal{L}_{{\chi}^0\tilde{U}u}=-\frac{i}{6}{\bar{\chi}^0_{i}}\Big\{\Big[\sqrt{2}(g_{1}N_{1i}+3g_{2}N_{2i}
+g_{YX}N_{5i})Z^{\tilde{U}^*}_{jk}+6N_{4i}Y^j_{u}Z^{\tilde{U}^*}_{(3+j)k}\Big]P_L
\nonumber\\&&\hspace{1.6cm}+\Big[6Y^j_{u}Z^{\tilde{U}^*}_{jk}
N^*_{4i}-\sqrt{2}Z^{\tilde{U}^*}_{(3+j)k}\Big((4g_{YX}+3g_{X})N^*_{5i}+4g_{1}N^*_{1i}\Big)\Big]P_R\Big\}u_{j}\tilde{U}^*_{k},
\\&&\mathcal{L}_{\chi^\pm\tilde{U}d}=\sum_{a=1}^3
\bar{d}_{i}\Big\{U^*_{j2}Z^{\tilde{U}^*}_{ka}Y^a_{d}P_L+
\Big[Y^a_{u}Z^{\tilde{U}*}_{k(3+a)}V_{j2}-g_2
Z^{\tilde{U}^*}_{ka}V_{j1}\Big]P_{R}\Big\}{\chi^\pm_{i}}\tilde{U}^*_{k},
\\&&\mathcal{L}_{\chi^\pm\tilde{D}u}=
\sum_{a=1}^3\bar{\chi}^\pm_{i}\Big\{\Big[U^*_{i2}Y^a_{d}Z^{\tilde{D}}_{k(3+a)}-g_2U^*_{i1}
Z^{\tilde{D}}_{ka}\Big]P_L+Y^a_{u}Z^{\tilde{D}*}_{ka}
Y^a*_{u}V_{i2}P_{R}\Big\}u_{j}\tilde{D}^*_{k}.
\end{eqnarray}
\section { Observation quantity and analysis result}
\subsection{Observables}
In this chapter, we'll introduce the observables $(\frac{R_{J/\psi}}
{R^{SM}_{J/\psi}}$, $\frac{R_{{D}_s}}{R^{SM}_{D_s}},
\frac{R_{{D^*}_s}}{R^{SM}_{D^*_s}},\frac{R_{{\Lambda}_c}}{R^{SM}_{\Lambda_c}}$)  and how to calculate them.When the energy of physical processes is far below the electroweak scale of $246$ GeV, heavy particles from both
 the SM and new-physics scenarios in the $U(1)_X$SSM, and various superparticles-do not directly take part in low-energy reactions. Their contributions can be treated in an effective way, yielding a universal effective
 theory to describe charged current decay processes such as $b\rightarrow{c~\tau~\nu}$\cite{20}.

  Only left-handed neutrinos exist in the SM, so all operators carry $P_L$ projector on neutrino spinors. The full dimension-6 effective Lagrangian reads:
\begin{eqnarray}
&&\mathcal{L}^LE_{b\rightarrow{c\tau\nu}}=-\frac{4G_{F}V_{cb}}{\sqrt{2}}
\Big[(C^{\tau}_{V_{L}}+C^{\tau}_{V_{L}}){O}^{\tau}_{V_{L}}
+C^{\tau}_{V_{R}}O^{\tau}_{V_{R}}+C^{\tau}_{S_{L}}O^{\tau}_{S_{L}}
+C^{\tau}_{S_{R}}O^{\tau}_{S_{R}}+C^{\tau}_{T}O^{\tau}_{T}\Big].
\end{eqnarray}

Here $G_F$ is Fermi constant, and $V_{cb}$
is the CKM \cite{34}matrix element for $b\rightarrow{c~\tau~\nu}$ mixing, and five independent Lorentz-invariant four-fermion operators are defined as:
\begin{eqnarray}
&&{O}^{\tau}_{V_{L}}=[\bar{c}\gamma^{\mu}P_{L}b][\bar{\tau}\gamma_{\mu}P_{L}\nu],
\\&&O^{\tau}_{V_{R}}=[\bar{c}\gamma^{\mu}P_{R}b][\bar{\tau}\gamma_{\mu}P_{L}\nu],
\\&&O^{\tau}_{S_{L}}=[\bar{c}P_{L}b][\bar{\tau}P_{L}\nu],
\\&&O^{\tau}_{S_{R}}=[\bar{c}P_{R}b][\bar{\tau}P_{L}\nu],
\\&&O^{\tau}_{T}=[\bar{c}\sigma^{\mu\nu}b][\bar{\tau}\sigma_{\mu\nu}P_{L}\nu].
\end{eqnarray}

The SM only generates the left-handed vector operator ${O}^{\tau}_{V_{L}}$ via off shell $W^\pm$ exchange at tree level, giving $C_{V_L}^\tau|_{SM}=1$. All other Wilson coefficients vanish in the SM: $
C^{SM}_{V_{R}}=C^{SM}_{S_{L}}=C^{SM}_{S_{R}}=C^{SM}_{T}=0$. BSM effects shift coefficients via corrections $\delta C_{V_L}^\tau$ and non-zero scalar, right-handed vector, tensor coefficients.

SM does not have $O^{\tau}_{V_{R}}$ at tree level through $W^{\pm}$. In the $U(1)_XSSM$, we take into account the one loop diagrams and    $\delta C_{V_L}^\tau, C_{V_R}^\tau,~ C^{\tau}_{S_{R}},~
C^{\tau}_{V_{R}},~C^{\tau}_T$ are all considered.
All four observables
$R_{J/\psi},R_{D_s},R_{D_s^*},R_{\Lambda_c}$ possess distinct form-factor overlaps with scalar, tensor, vector operators.
We reproduce the full analytical formulae :
$\frac{R_{J/\psi}}
{R^{SM}_{J/\psi}}$, $\frac{R_{{D}_s}}{R^{SM}_{D_s}}$,
$\frac{R_{D^*_s}}{R^{SM}_{D^*_s}}$ and $\frac{R_{{\Lambda}_c}}{R^{SM}_{\Lambda_c}}$ are in Ref.\cite{6} as follows:
\begin{eqnarray}
&&\frac{R_{J/\psi}}{R^{SM}_{J/\psi}}=1.0+
\Re\Big(0.12C^{\tau}_{S_{L}}+0.034|C^{\tau}_{S_{L}}|^{2}
-0.12C^{\tau}_{S_{R}}-0.068C^{\tau}_{S_{L}}C^{\tau*}_{S_{R}}
\nonumber\\&&\hspace{1.3cm}+0.034|C^{\tau}_{S_{R}}|^2-5.3C^{\tau}_{T}+13|C^{\tau}_{T}|^2
-1.9C^{\tau}_{V_{R}}-0.12C^{\tau}_{S_{L}}C^{\tau*}_{V_{R}}
\nonumber\\&&\hspace{1.3cm}+0.12C^{\tau}_{S_{R}}C^{\tau*}_{V_{R}}+5.8C^{\tau}_{T}C^{\tau*}_{V_{R}}
+1.0|C^{\tau}_{V_{R}}|^2+2.0\delta{C^{\tau}_{V_{L}}}+0.12C^{\tau}_{S_{L}}\delta{C^{\tau*}_{V_{L}}}
\nonumber\\&&\hspace{1.3cm}-0.12C^{\tau}_{S_{R}}\delta{C^{\tau*}_{V_{L}}}
-5.3C^{\tau}_{T}\delta{C^{\tau*}_{V_{L}}}-1.9C^{\tau}_{V_{R}}\delta{C^{\tau*}_{V_{L}}}
+1.0|\delta{C^{\tau}_{V_{L}}}|^2\Big),\label{o1}
\\
&&\frac{R_{{D}_s}}{R^{SM}_{D_s}}=1.0
+\Re\Big(1.6C^{\tau}_{S_{L}}
+1.2|C^{\tau}_{S_{L}}|^{2}+1.6C^{\tau}_{S_{R}}+2.4C^{\tau}_{S_{L}}C^{\tau}_{S_{R}^*}
+1.2|C^{\tau}_{S_{R}}|^2\nonumber\\&&\hspace{1.3cm}+1.4C^{\tau}_{T}+1.4|C^{\tau}_{T}|^2
+2.0C^{\tau}_{V_{R}}+1.6C^{\tau}_{S_{L}}C^{\tau*}_{V_{R}}
+1.6C^{\tau}_{S_{R}}C^{\tau*}_{V_{R}}
+1.4C^{\tau}_{T}C^{\tau*}_{V_{R}}\nonumber\\&&\hspace{1.3cm}
+1.0|C^{\tau}_{V_{R}}|^2+2.0\delta{C^{\tau}_{V_{L}}}
+1.6C^{\tau}_{S_{L}}\delta{C^{\tau*}_{V_{L}}}
+1.6C^{\tau}_{S_{R}}\delta{C^{\tau*}_{V_{L}}}
+1.4C^{\tau}_{T}\delta{C^{\tau*}_{V_{L}}}\nonumber\\&&\hspace{1.3cm}+2.0C^{\tau}_{V_{R}}
\delta{C^{\tau*}_{V_{L}}}+1.0|\delta{C^{\tau}_{V_{L}}}|^2\Big),
\label{o2}
\\
&&\frac{R_{{D}_s}^*}{R^{SM}_{D^*_s}}=1.0+\Re\Big(0.085C^{\tau}_{S_{L}}
+0.026|C^{\tau}_{S_{L}}|^{2}-0.085C^{\tau}_{S_{R}}-0.052C^{\tau}_{S_{L}}C^{\tau*}_{S_{R}}
\nonumber\\&&\hspace{1.3cm}+0.026|C^{\tau}_{S_{R}}|^2-4.6C^{\tau}_{T}+15|C^{\tau}_{T}|^2
-1.8C^{\tau}_{V_{R}}-0.085C^{\tau}_{S_{L}}C^{\tau*}_{V_{R}}
\nonumber\\&&\hspace{1.3cm}+0.085C^{\tau}_{S_{R}}C^{\tau*}_{V_{R}}
+6.4C^{\tau}_{T}C^{\tau*}_{V_{R}}
+1.0|C^{\tau}_{V_{R}}|^2+2.0\delta{C^{\tau}_{V_{L}}}+0.085C^{\tau}_{S_{L}}\delta{C^{\tau*}_{V_{L}}}
\nonumber\\&&\hspace{1.3cm}-0.085C^{\tau}_{S_{R}}\delta{C^{\tau*}_{V_{L}}}
-4.6C^{\tau}_{T}\delta{C^{\tau*}_{V_{L}}}-1.8C^{\tau}_{V_{R}}
\delta{C^{\tau*}_{V_{L}}}+1.0|\delta{C^{\tau}_{V_{L}}}|^2\Big).\label{o3}
\\
&&\frac{R_{{\Lambda}_c}}{R^{SM}_{\Lambda_c}}=1.0+\Re\Big(0.39C^{\tau}_{S_{L}}
+0.34|C^{\tau}_{S_{L}}|^{2}-0.49C^{\tau}_{S_{R}}
-0.61C^{\tau}_{S_{L}}C^{\tau*}_{S_{R}}\nonumber\\&&\hspace{1.3cm}+0.34|C^{\tau}_{S_{R}}|^2+1.1C^{\tau}_{T}+12|C^{\tau}_{T}|^2-0.71C^{\tau}_{V_{R}}+0.49C^{\tau}_{S_{L}}C^{\tau*}_{V_{R}}
\nonumber\\&&\hspace{1.3cm}+0.39C^{\tau}_{S_{R}}C^{\tau*}_{V_{R}}-1.7C^{\tau}_{T}C^{\tau*}_{V_{R}}+1.0|C^{\tau}_{V_{R}}|^2+2.0\delta{C^{\tau}_{V_{L}}}
+0.39C^{\tau}_{S_{L}}\delta{C^{\tau*}_{V_{L}}}
\nonumber\\&&\hspace{1.3cm}+0.49C^{\tau}_{S_{R}}\delta{C^{\tau*}_{V_{L}}}
+1.1C^{\tau}_{T}\delta{C^{\tau*}_{V_{L}}}
-0.71C^{\tau}_{V_{R}}\delta{C^{\tau*}_{V_{L}}}
+1.0|\delta{C^{\tau}_{V_{L}}}|^2\Big).
\label{o4}
\end{eqnarray}
The tensor operator $C_T^\tau$ dominates the quadratic contributions here. Accordingly, $R_{D_s^*}$ serves as the primary discriminator for tensor type new physics among the four observables.

\subsection{Analysis result}
\begin{figure}[ht]
\subfigure[]{
\setlength{\unitlength}{5.0mm}
\includegraphics[width=1.5in]{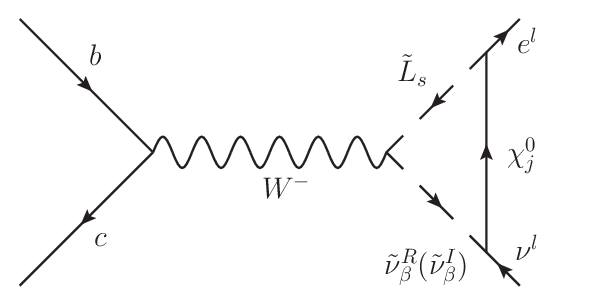}
\label{Fig1}
}
\subfigure[]{
\setlength{\unitlength}{5.0mm}
\includegraphics[width=1.5in]{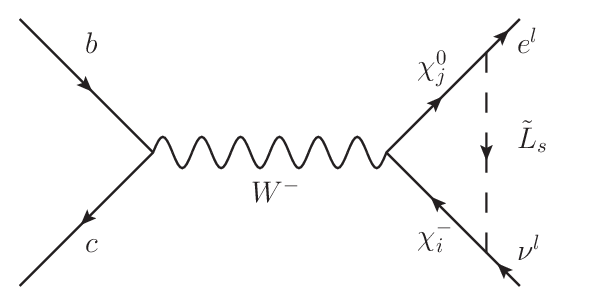}
\label{Fig2}
}
\subfigure[]{
\setlength{\unitlength}{5.0mm}
\includegraphics[width=1.5in]{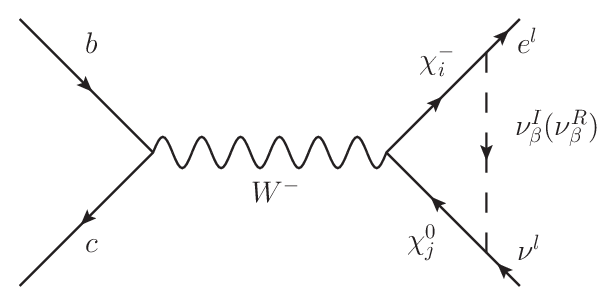}
\label{Fig3}
}
\subfigure[]{
\setlength{\unitlength}{5.0mm}
\includegraphics[width=1.5in]{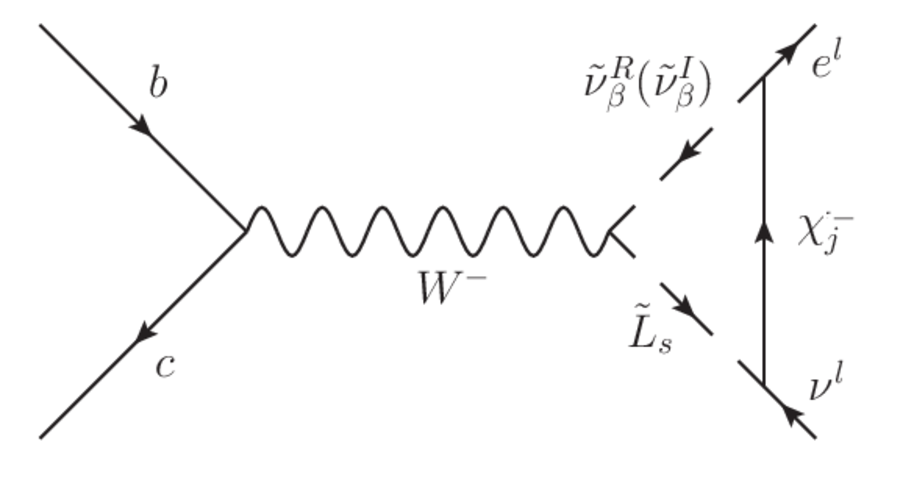}
\label{Fig4}
}
\subfigure[]{
\setlength{\unitlength}{5.0mm}
\includegraphics[width=1.5in]{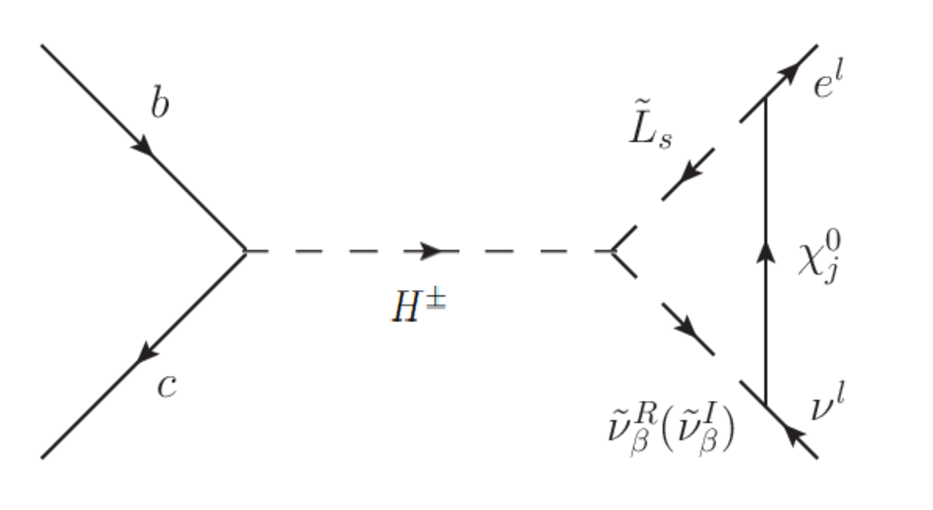}
\label{Fig5}
}
\subfigure[]{
\setlength{\unitlength}{5.0mm}
\includegraphics[width=1.5in]{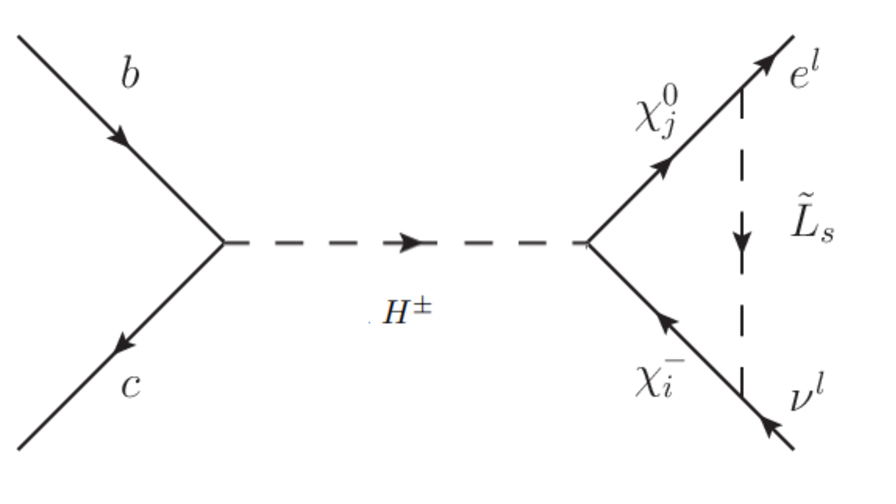}
\label{Fig6}
}
\subfigure[]{
\setlength{\unitlength}{5.0mm}
\includegraphics[width=1.5in]{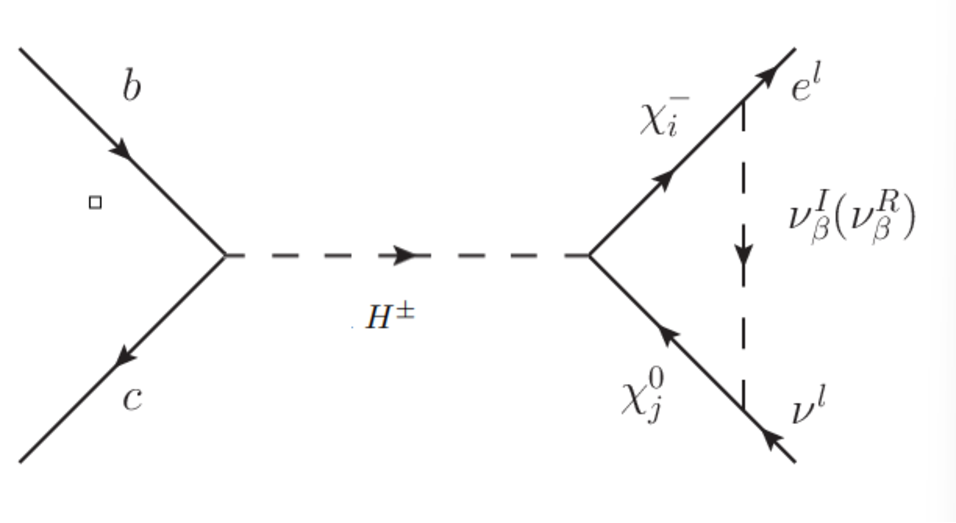}
\label{Fig7}
}
\subfigure[]{
\setlength{\unitlength}{5.0mm}
\includegraphics[width=1.5in]{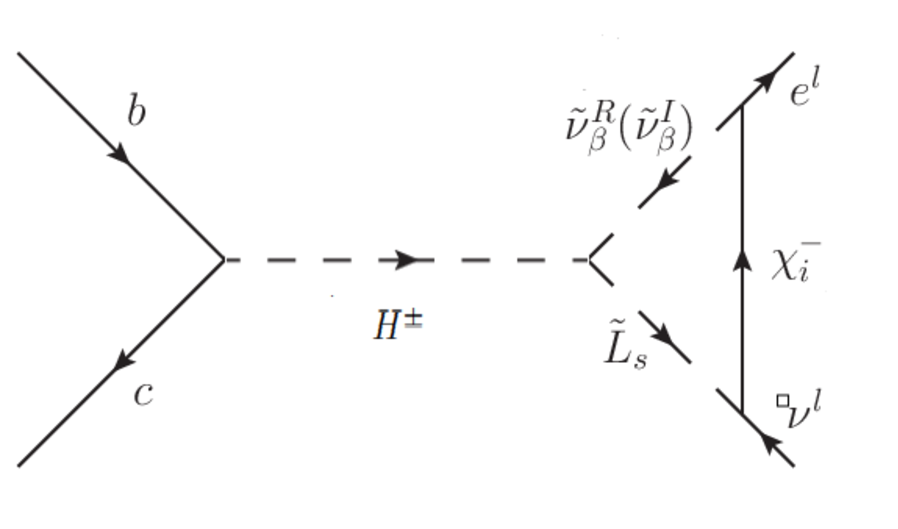}
\label{Fig8}
}
\subfigure[]{
\setlength{\unitlength}{5.0mm}
\includegraphics[width=1.5in]{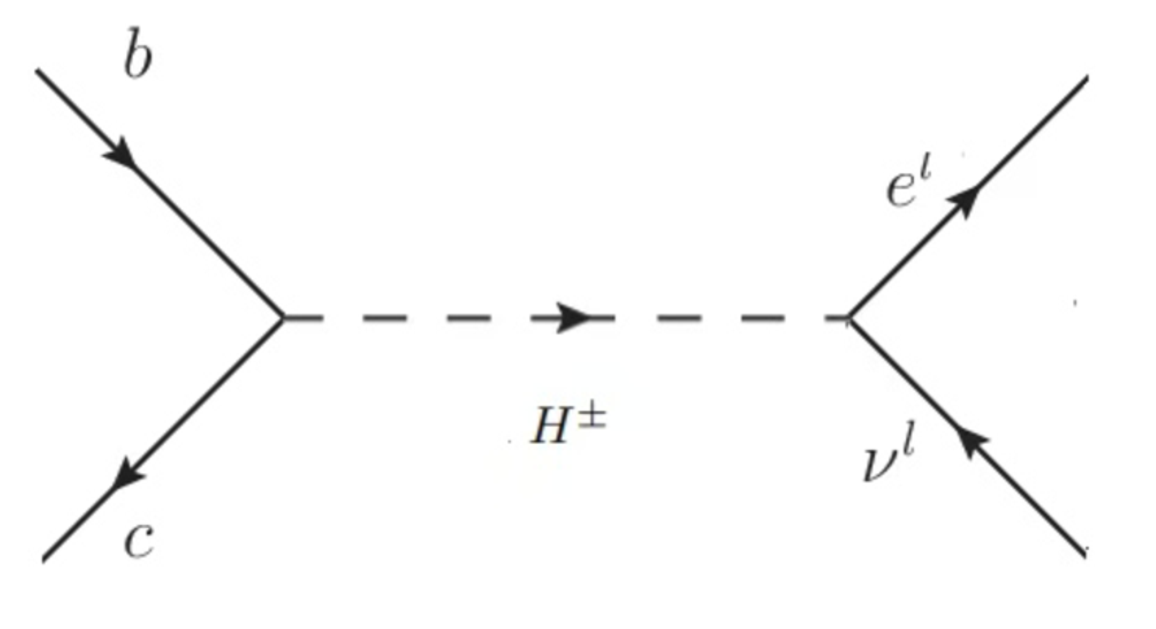}
\label{Fig9}
}
\caption{The penguin-type feynman diagrams for the $b\rightarrow c l \nu$ process.}\label{N1}
\end{figure}
Fig.\ref{N1} and Fig.\ref{N2} are the Feynman diagrams for $b\rightarrow cl\nu$ that we use to calculate the observables under the $U1_X$SSM, including penguin-type and box-type Feynman diagrams.
To save space in the text, here we take Fig.\ref{N1} (a) as an example and obtain the specific expression for the Feynman amplitude as follows:
\begin{eqnarray}
\mathcal{C}_{VL}^{\ell}&=&-\frac{1}{\sqrt{2}G_{F}V_{cb}}
\sum_{\beta,s=1}^{6}\sum_{j=1}^{8}\frac{\mathcal{B}_{1}^{\ell sj}\mathcal{A}_{2}^{\beta\ell j}\mathcal{A}_{3}^{\beta
s}\mathcal{A}_{4}}{m_{W}^{2}}\frac{1}{64\pi^{2}}F_{1}(x_{\chi^0_j},x_{\tilde{L}_{s}},x_{\tilde{\nu}^{I}_{\beta}} ),\nonumber\\&&\hspace{-1.5cm}\mathcal{C}_{AL}^{\ell}~=-\mathcal{C}_{VL}^{\ell}.
\label{Ca}
\end{eqnarray}

Here we  list the coupling vertices used in the simplified version.
\begin{eqnarray}
&&\mathcal{A}_{3}^{\beta s}=-\frac{1}{2}g_2\sum_{a=1}^3Z^{E*}_{sa}Z^{R*}_{\beta a},\hspace{1cm}\mathcal{A}_{4}=-\frac{1}{\sqrt{2}}g_2V_{cb},\nonumber\\&&\hspace{0cm}
\mathcal{B}_{1}^{\ell sj}=\frac{1}{\sqrt{2}}(g_1N_{j1}+g_2N_{j2}+g_{YX}N_{j5})Z^{E*}_{s \ell}
-N_{j3}Y^\ell_lZ^{E*}_{s(3+\ell)},\nonumber\\&&\hspace{0cm}
\mathcal{A}_{2}^{\beta\ell j}=(g_2N^*_{j2}-g_{YX}N^*_{j5}-g_1N^*_{j1}) \sum_{a=1}^3Z^{I*}_{\beta a}U_{\ell a}^{V*},\nonumber\\&&\hspace{0cm}\nonumber
\end{eqnarray}
To calculate this diagram, we need to use the following one loop  function:
\begin{eqnarray}
&&F_{1}(x_{1},x_{2},x_{3})=\frac{x_{1}\ln x_{1}}{(x_{1}-x_{2})(x_{1}-x_{3})}+\frac{x_{2}\ln x_{2}}{(x_{2}-x_{1})(x_{2}-x_{3})}+\frac{x_{3}\ln x_{3}}{(x_{3}-x_{1})(x_{3}-x_{2})},
\nonumber\\&&\hspace{0cm}
\end{eqnarray}
Here $x_i=\frac{m_i^2}{M^2_{SUSY}}$ with $M_{SUSY}$ denoting the supersymmetric mass scale.
According to the above procedure, we can get the coefficients of all the diagrams in Fig.~\ref{N1}.
The Feynman diagram we used as an example is divergent, so dimensional regularization\cite{35} is needed to handle the divergent terms. To get a finite result, the divergent terms are cancelled by the modified minimal
subtraction ($\overline{{MS}} $) scheme.

\subsection{Box-type Feynman diagrams}
\begin{figure}[ht]
\centering
\subfigure[]{
\setlength{\unitlength}{5.0mm}
\includegraphics[width=1.4in]{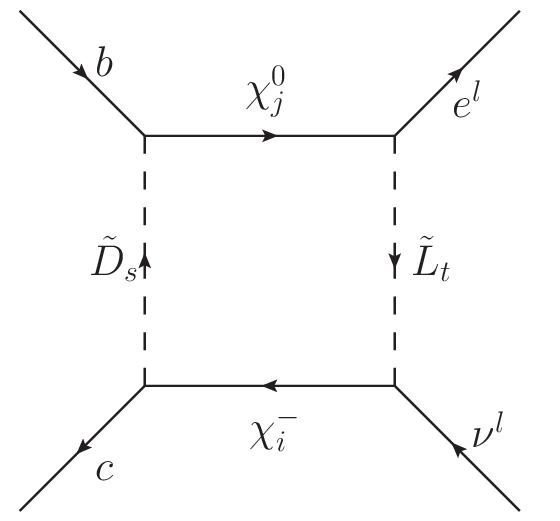}
\label{Fig10}
}
\subfigure[]{
\setlength{\unitlength}{5.0mm}
\includegraphics[width=1.4in]{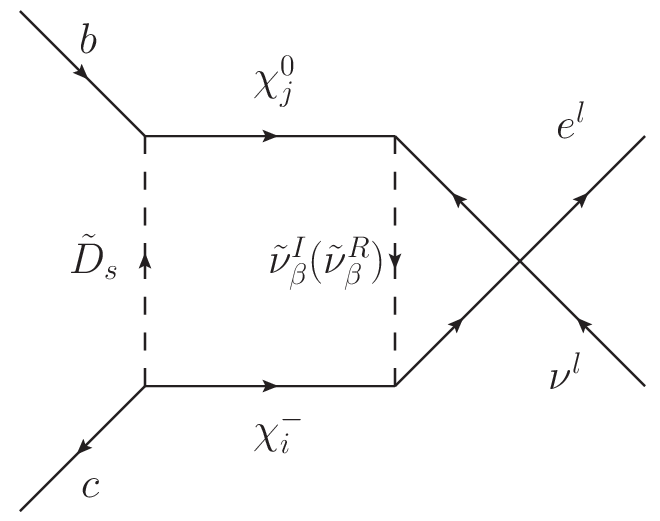}
\label{Fig11}
}
\caption{The box-type feynman diagrams for the process $b\rightarrow c l \nu$.}\label{N2}
\end{figure}
Unlike the Feynman diagrams in the previous picture, the feynman diagrams in Fig.\ref{N2} converge, and the used method is the same as above.
Taking Fig.~\ref{N2} (a) as an example, the corresponding WCs are given as follows:
\begin{eqnarray}
&&\mathcal{C}_{VL}=
\sum_{j=1}^{8}\sum_{k_1,k_2=1}^{6}\sum_{j_1=1}^{2}
\frac{1}{4\sqrt{2}G_{F}V_{cb}}{\mathcal{A}_{L}}
\mathcal{B}_{R}\mathcal{C}_{R}\mathcal{D}_{L} F_{2}(x_{\tilde{L}_{k_1}},
x_{\chi^0_{j}},x_{\tilde{Q}_{k_2}},x_{\chi^\pm_{j_1}}),
\label{CVL}
\\&&
\mathcal{C}_{VR}=\sum_{j=1}^{8}\sum_{k_1,k_2=1}^{6}\sum_{j_1=1}^{2}
-\frac{1}{2\sqrt{2}G_{F}V_{cb}}m_{\chi^0_{j}}\sqrt{x_{\chi^\pm_{j_1}}}{\mathcal{A}_{R}}\mathcal{B}_{R}\mathcal{C}_{L}\mathcal{D}_{L}
 F_{3}(x_{\tilde{L}_{k_1}},
x_{\chi^0_{j}},x_{\tilde{Q}_{k_2}},x_{\chi^\pm_{j_1}}),
\label{CVR}
\\&&\mathcal{C}_{SL}=\sum_{j=1}^{8}\sum_{k_1,k_2=1}^{6}\sum_{j_1=1}^{2}
-\frac{1}{\sqrt{2}G_{F}V_{cb}}m_{\chi^0_{j}}\sqrt{x_{\chi^\pm_{j_1}}}(\frac{1}{8}{\mathcal{A}_{R}}\mathcal{B}_{R}\mathcal{C}_{L}\mathcal{D}_{L}
\nonumber\\&&\hspace{1.3cm}-\frac{3}{8}{\mathcal{A}_{L}}\mathcal{B}_{L}\mathcal{C}_{L}\mathcal{D}_{L})
 F_{3}(x_{\tilde{L}_{k_1}},
x_{\chi^0_{j}},x_{\tilde{Q}_{k_2}},x_{\chi^\pm_{j_1}}),
\label{CSL}
\\&&
\mathcal{C}_{SR}=\sum_{j=1}^{8}\sum_{k_1,k_2=1}^{6}\sum_{j_1=1}^{2}
\Big(-\frac{1}{2}{\mathcal{A}_{R}}\mathcal{B}_{L}\mathcal{C}_{R}\mathcal{D}_{L} F_{2}(x_{\tilde{L}_{k_1}},
x_{\chi^0_{j}},x_{\tilde{Q}_{k_2}},x_{\chi^\pm_{j_1}}) \nonumber\\&&\hspace{1.3cm}
-m_{\chi^0_{j}}\sqrt{x_{\chi^\pm_{i}}}\frac{1}{8}
({\mathcal{A}_{R}}\mathcal{B}_{R}
+{\mathcal{A}_{L}}\mathcal{B}_{L}
)\mathcal{C}_{L}\mathcal{D}_{L}F_{3}(x_{\tilde{L}_{k_1}},
x_{\chi^0_{j}},x_{\tilde{Q}_{k_2}},x_{\chi^\pm_{j_1}})\Big )/(\sqrt{2}G_{F}V_{cb}),
\label{CSR}
\end{eqnarray}
and all the other WCs  vanish. In Eqs. (\ref{CVL}-\ref{CSR}),
\begin{eqnarray}
&&\mathcal{A}_{L}=-\frac{1}{6}(\sqrt{2}g_{1}N^*_{j,1}Z^{D}_{k1,i}-3\sqrt{2}g_{2}N^*_{j2}Z^{D}_{k1,i}+\sqrt{2}g_{YX}N^*_{j,5}Z^{D}_{ki}+6\sqrt{2}g_{1}N^*_{j,3}Z^{D}_{k1,i+3}Y_{d,i})\nonumber\\&&\hspace{0cm}
\mathcal{B}_{L}=(-\frac{1}{\sqrt{2}}Z^{E,*}_{k2,3+i}(2g_{1}N^*_{j,1}+(2g_{YX}+g_{X})N^*_{j,5}))-N^*_{j,3}Z^{E,*}_{k2,i}Y_{e,i},\nonumber\\&&\hspace{0cm}
\mathcal{C}_{L}=V^*_{j1,2}Z^{D}_{k1,i}Y_{u,i},\nonumber\\&&\hspace{0cm}
\mathcal{D}_{L}=-g_{2}U_{j1,1}\sum_{a=1}^{3}{U_{i,a}}^{V*}Z_{k2,a}^{E}+U^*_{j1,2}\sum_{a=1}^{3}{U_{i,a}}^{V*}Z_{k2,a+3}^{E}Y_{e,a},\nonumber\\&&\hspace{0cm}
\mathcal{A}_{R}=-\frac{1}{6}\Big(6Y^*_{d,i}Z^{D}_{k1,i}N_{j,3}
+\sqrt{2}Z^{D}_{k1,i+3}(2g_{1}N^*_{j,1}+(2g_{YX}+3g_X)N^*_{j,5})\Big)
,\nonumber\\&&\hspace{0cm}
\mathcal{B}_{R}=\frac{1}{\sqrt{2}}Z^{E,*}_{k2,i}(g_{1}N_{j,1}+g_{2}N_{j,2}+g_{YX}N_{j,5})
-Z^{E,*}_{k2,3+i}Y_{e,i},\nonumber\\&&\hspace{0cm}
\mathcal{C}_{R}=-g_{2}Z^{D*}_{k1,i}+U_{j1,1}+U_{j1,2}Y^*_{d,i}Z^{D*}_{k1,i+3}.
\end{eqnarray}

The formulae $F_{2}(x_{1},x_{2},x_{3},x_{4})$ and $F_{3}(x_{1},x_{2},x_{3},x_{4})$ are given as follows:
 \begin{eqnarray}
&&F_{2}(x_{1},x_{2},x_{3},x_{4})=\frac{x_{1}^{2}\ln x_{1}}{(x_{1}-x_{2})(x_{1}-x_{3})(x_{1}-x_{4})}+\frac{x_{2}^{2}\ln x_{2}}{(x_{2}-x_{1})(x_{2}-x_{3})(x_{2}-x_{4})}\nonumber\\&&\hspace{3.3cm}
+\frac{x_{3}^{2}\ln x_{3}}{(x_{3}-x_{1})(x_{3}-x_{2})(x_{3}-x_{4})}+\frac{x_{4}^{2}\ln x_{4}}{(x_{4}-x_{1})(x_{4}-x_{2})(x_{4}-x_{3})},\nonumber\\&&\hspace{0cm}
F_{3}(x_{1},x_{2},x_{3},x_{4})=\frac{x_{1}\ln x_{1}}{(x_{1}-x_{2})(x_{1}-x_{3})(x_{1}-x_{4})}+\frac{x_{2}\ln x_{2}}{(x_{2}-x_{1})(x_{2}-x_{3})(x_{2}-x_{4})}\nonumber\\&&\hspace{3.3cm}
+\frac{x_{3}\ln x_{3}}{(x_{3}-x_{1})(x_{3}-x_{2})(x_{3}-x_{4})}+\frac{x_{4}\ln x_{4}}{(x_{4}-x_{1})(x_{4}-x_{2})(x_{4}-x_{3})}.
\end{eqnarray}
We process all the Feynman diagrams in the figure using the methods mentioned above and further simplify them.
By deriving the amplitude expressions for  the Feynman diagrams, we can obtain  the non-zero Wilson coefficients, which then allows us to calculate the values of these observables in Eqs.(\ref{o1}-\ref{o4}).
We then compare them with the Wilson coefficients in the SM, and calculate the complex expressions numerically to obtain the results.

\section { Numerical analysis}
In the numerical analysis of this chapter, we take into account the following experimental constraints:

   1. The mass of the lightest CP-even Higgs boson mh agrees with the experimental result $m_h =125.13\pm0.11$ GeV\cite{36}.

   2. The new angle $\beta_\eta$ is constrained by LHC as $\tan \beta_\eta< 1.5$ \cite{37}.

   3. The constraints for the particle masses accord to the PDG~\cite{36} data,and the concrete contents are the following.
   The neutralino mass is limited to more than $116~\rm{GeV}$,
   The masses of chargino, slepton and  squark are at the $\rm{TeV}$order\cite{38}.

 We use images to visualise how variables affect the results, and below we'll use one-dimensional and two-dimensional charts to show our analysis of the numbers.
\subsection{One-dimensional image}
To draw the linear graph, we use the following parameters as variables
\begin{eqnarray}
l_W,~  (T_{\nu})_{ii}, ~(T_{e})_{ii}, ~(M_{E})_{ii},
~M_2, ~(Y_X)_{ii}, ~i=1,~2,~3.
\end{eqnarray}
We use images to visualize the effects of variables on the results. Under the premise of meeting the above experimental constraints, we use the parameters as follows:
\begin{eqnarray}
&&\tan\beta_{\eta}=1.05,~~~\lambda_C=-0.17,~~~\lambda_H=-0.48,~~~\mu=1~{\rm TeV},
\nonumber\\&& M_{BB'}=0.1~{\rm TeV},~~ M_{BL}=1~{\rm TeV},~~M_S=0.5~{\rm TeV},~~\kappa=-1,
\nonumber\\&&({M_\nu}^2)_{11}=({M_\nu}^2)_{22}=({M_\nu}^2)_{33}=0.3~{\rm TeV}^2,
~~({T_x})_{ii}=11~{\rm TeV},\nonumber\\&&~
({m_{\tilde{D}}^2})_{ii}=2.8^2~{\rm{TeV}^2},~~
({m_{\tilde{Q}}^2})_{ii}=({m_{\tilde{U}}^2})_{ii}=3~{\rm{TeV}^2},~ v_S=7~{\rm TeV}.
\end{eqnarray}
1. {The effects of parameters on $\frac{R_{J/\psi}}{R^{SM}_{J/\psi}}$}\\
\begin{figure}[ht]
\centering
\subfigure[]{
\setlength{\unitlength}{5.0mm}
\includegraphics[width=2.5in]{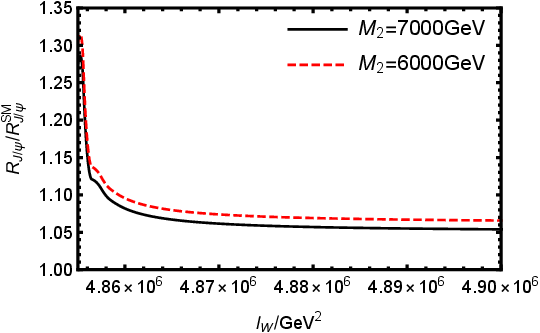}
\label{Fig12}
}
\subfigure[]{
\setlength{\unitlength}{5.0mm}
\includegraphics[width=2.5in]{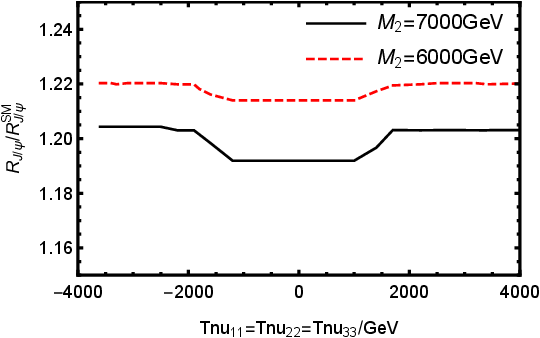}
\label{Fig13}
}
\caption{The effects of $l_W$ and  $(T_{\nu})_{ii}$ on the ratio $(\frac{R_{J/\psi}}{R^{SM}_{J/\psi}})$.} {\label {fig3}}
\end{figure}

  We first study the effect of the parameter $l_W$  on the observed quantity $(\frac{R_{J/\psi}}{R^{SM}_{J/\psi}})$. $l_{W}$ affects the mass matrixes of
  neutral Higgs and charged Higgs.
  Therefore, it is an important parameter to the process $b\rightarrow c l\nu$.
  For $l_W$, the range we take is $(4.85\times 10^6 -4.9\times 10^6~ {\rm GeV}^2)$, and when $l_W$ changes within this range, we plot the relationship between $l_W$ and $\frac{R_{J/\psi}}{R^{SM}_{J/\psi}}$  in
  Fig.\ref{fig3} (a).
  The dashed line represents $M_2$=6000GeV, and the solid line represents $M_2$=7000GeV.
  $\frac{R_{J/\psi}}{R^{SM}_{J/\psi}}$ drops rapidly and then flattens out as $l_{W}$ increases. The parameter $l_W$ is dominant for the new-physics contributions: small $l_W$ yields large corrections, while larger $l_W$
  suppresses such effects.
  The plateau at large $l_W$ signals residual new-physics contributions insensitive to $l_W$.
  Large $M_2$ leads to heavy chargino mass and neutralino mass.
  A smaller particle mass induces stronger modification. So the dashed curve lies above the solid one.


  The Fig.\ref{fig3} (b) illustrates the variation of the observable ratio with the trilinear coupling parameter $(T_\nu)_{ii}$ for sneutrino.
   $(T_\nu)_{ii}$ is an effective-interaction parameter governing the sneutrino sector, which contributes to the results of the $b\rightarrow c\tau\nu$ decay through the sneutrino vertexes relating with
   chargino(neutralino).
   Nevertheless, this parameter influences the results weakly for the ratio $\frac{R_{J/\psi}}{R^{SM}_{J/\psi}}>1$.
   Even when this parameter takes large positive or negative values, the resulting modification of the decay amplitude remains small. Compared with $l_W$, $(T_\nu)_{ii}$ is an insensitive parameter.
   Within the parameter range from $-4000~{\rm GeV}$ to $4000~{\rm GeV}$, the curve exhibits only minor variations: a shallow dip appears near the point $(T_\nu)_{ii}=0 ~{\rm GeV}$, and the ratio rises slightly as
   $|(T_\nu)_{ii}|\geq ~2000 {\rm GeV}$. Therefore, the overall "dip in the middle and rise on both sides" shape of the curve is preserved. The curve for $M_2=6000~{\rm GeV}$ lies globally above that for $M_2=7000~{\rm
   GeV}$. A smaller particle mass scales up the new-physics contributions proportionally, so the dashed curve sits above the solid one.


 2.{The effects of parameters on $\frac{R_{{D}_s}}{R^{SM}_{D_s}}$ and $\frac{R_{{D^*}_s}}{R^{SM}_{D^*_s}}$}

\begin{figure}[ht]
\centering
\subfigure[]{
\setlength{\unitlength}{5.0mm}
\includegraphics[width=2.5in]{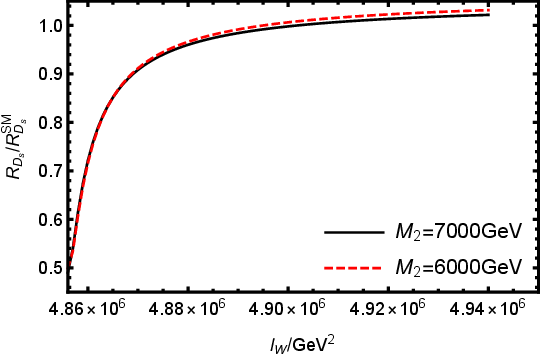}
\label{Fig14}
}
\subfigure[]{
\setlength{\unitlength}{5.0mm}
\includegraphics[width=2.5in]{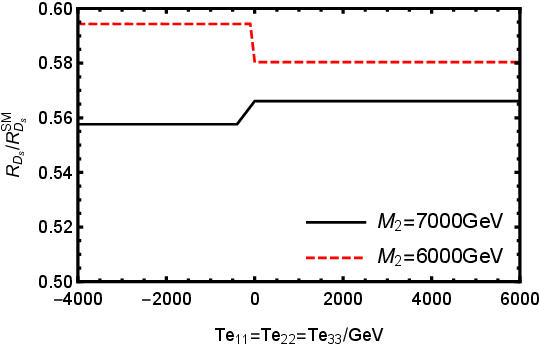}
\label{Fig15}
}
\caption{The effects of $l_W$ and  $(T_{e})_{ii}$ on the observed quantity $\frac{R_{{D}_s}}{R^{SM}_{D_s}}$.} {\label {fig4}}.
\end{figure}

 There aren't exact experimental values for $\frac{R_{{D}_s}}{R^{SM}_{D_s}}$ and
$\frac{R_{{D^*}_s}}{R^{SM}_{D^*_s}}$, but we'll still draw a few graphs to see how the parameters $l_W$, $(T_{e})_{ii}$ and $(M_{E})_{ii}$ affect them in the Fig.\ref{fig4} .
Similar as Fig.\ref{fig3}, we still use a black solid line to show the variation trend of $\frac{R_{{D}_s}}{R^{SM}_{D_s}}$ and $\frac{R_{{D^*}_s}}{R^{SM}_{D^*_s}}$ , when $M_2$ is 7000 GeV, and a red dashed line to show the
trend with $M_2= 6000$ GeV.

Fig.\ref{fig4} (a) shows the evolution of the observable ratio $\frac{R_{{D}_s}}{R^{SM}_{D_s}}$ as a function of the parameter $l_W$.
 The ratio $\frac{R_{J/\psi}}{R^{SM}_{J/\psi}}$ is larger than 1.
 On the contrary, $\frac{R_{{D}_s}}{R^{SM}_{D_s}} <1$ as shown in the Fig.\ref{fig4}(a).
 In contrast to the enhancement effect observed for $R_{J/\psi}$, destructive interference between the new physics and SM amplitudes occurs here.
 For small $l_W$, the destructive interference is substantial and the observable is strongly suppressed. As $l_W$ increases, the corresponding new-physics contribution is suppressed, and the ratio gradually rises toward the
 SM prediction.
 Residual new-physics effects persist even at large $l_W$, so the curve cannot fully reach unity. Comparing the two sets of $M_2$ parameters, the both curves lie very close to each other, indicating that $R_{D_s}$ is less
 sensitive to the heavy-particle mass $M_2$ than $R_{J/\psi}$.

 Fig.\ref{fig4} (b) shows the behavior of $\frac{R_{{D}_s}}{R^{SM}_{D_s}}$ as a function of the trilinear parameter $(T_{e})_{ii}$ of the slepton. The mass squared matrix of slepton has the parameter $(T_e)_{ii}$.
 So, the masses and mixing of slepton are affected by $(T_e)_{ii}$.
 The amplitude for $b\rightarrow c\tau\nu$ includes the slepton-chargino(neutralino) contribution.
  The curve exhibits step-like features: the observable undergoes a weak abrupt jump as $(T_e)_{ii}$ crosses zero from negative to positive values and remains stable in the negative- and positive-parameter regions,
  respectively. This indicates that the sign of $(T_e)_{ii}$ determines the interference nature of this effective interaction, rather than providing a continuous tuning of the interaction strength. For the two sets of $M_2$
  parameters, the jump directions are opposite, which demonstrates that $R_{D_s}$ is sensitive to the combined sign of $M_2$ and $(T_e)_{ii}$. Overall, the extent of influence from $(T_e)_{ii}$ on $R_{D_s}$ is much smaller
  than that of $l_W$, which implies that $(T_e)_{ii}$ affects the results mildly.
  Through correlations among different operators, $(T_e)_{ii}$ induces sub-leading corrections to the observable $R_{D_s}$.

\begin{figure}[ht]
\centering
\subfigure[]{
\setlength{\unitlength}{5.0mm}
\includegraphics[width=2.5in]{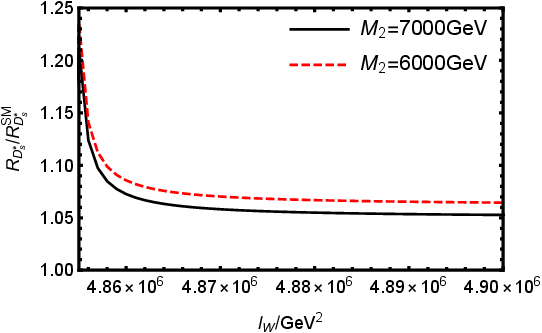}
\label{Fig16}
}
\subfigure[]{
\setlength{\unitlength}{5.0mm}
\includegraphics[width=2.5in]{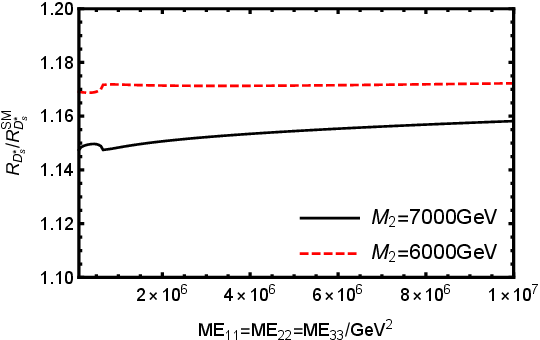}
\label{Fig17}
}
\caption{ The effects of $l_W$ and  $(M_{E})_{ii}$ on the observed quantity $\frac{R_{{D^*}_s}}{R^{SM}_{D^*_s}}$.} {\label {fig5}.}
\end{figure}

Fig.\ref{fig5}(a) shows the evolution of $\frac{R_{{D^*}_s}}{R^{SM}_{D^*_s}}$ as a function of the parameter $l_W$. Similar to $R_{J/\psi}$, this ratio is globally larger than unity, indicating constructive interference
between the new-physics and SM amplitudes, which enhances the $D_s^*$-decay observable.
The corrections are substantial for small $l_W$. As $l_W$ increases, the corresponding new-physics contribution is suppressed, and the ratio drops rapidly and saturates in the large-$l_W$ region. In contrast to the
destructive-interference behavior of the  $D_s$, the $D_s^*$ possesses a different helicity structure, leading to an opposite interference sign. This character is similar as that of $J/\psi$. The curve for $M_2=6000~{\rm
GeV}$ lies above that for $M_2=7000~{\rm GeV}$, consistent with the rule that a lighter particle induces larger corrections.

Fig.\ref{fig5}(b) illustrates the evolution of $\frac{R_{{D^*}_s}}{R^{SM}_{D^*_s}}$ as a function of the parameter $(M_E)_{ii}$.
The parameter $(M_E)_{ii}$ appears in the slepton mass squared matrix,
 mainly affect slepton masses and the couplings associated with
 slepton and chargino(neutralino).
 Its effects embody through the loop diagrams.
 As $(M_E)_{ii}$ increases, the ratio for the $D_s^*$ decay undergoes only slight modifications, indicating $(M_E)_{ii}$ as a non-sensitive parameter.
  The curve corresponding to $M_2=6000~{\rm GeV}$ sits globally higher
 than the line with $M_2=7000~{\rm GeV}$.

    3. {The effects of parameters on $\frac{R_{{\Lambda}_c}}{R^{SM}_{\Lambda_c}}$}
\begin{figure}[ht]
\centering
\subfigure[]{
\setlength{\unitlength}{5.0mm}
\includegraphics[width=2.5in]{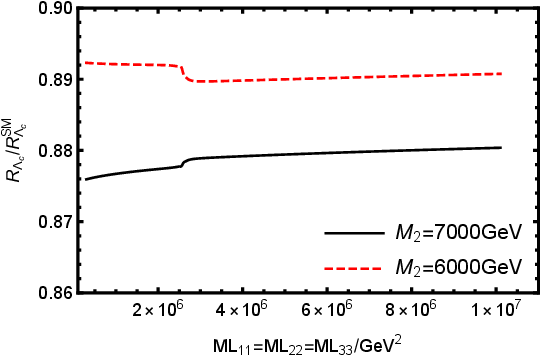}
\label{Fig18}
}
\subfigure[]{
\setlength{\unitlength}{5.0mm}
\includegraphics[width=2.5in]{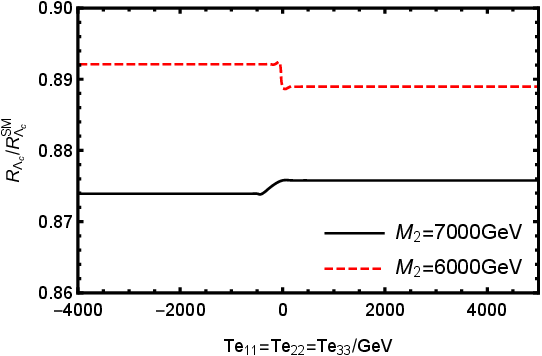}
\label{Fig19}
}
\caption{The effects of $(M_{L})_{ii}$ and  $(T_{e})_{ii}$ on the observed quantity $\frac{R_{{\Lambda}_c}}{R^{SM}_{\Lambda_c}}$.} {\label {fig6}.}
\end{figure}

  In Fig.\ref{fig6}(a), we analyze the evolution of the ratio $\frac{R_{{\Lambda}_c}}{R^{SM}_{\Lambda_c}}$
  with the parameter $(M_L)_{ii}$. The parameter $(M_L)_{ii}$ appears in the slepton, CP-even(odd) sneutrino mass squared matrixes. Therefore, slepton masses,
   snetrino masses and their related couplings are all affected by $(M_L)_{ii}$.
  The numerical results show that the curve of the ratio $\frac{R_{{\Lambda}_c}}{R^{SM}_{\Lambda_c}}$ only presents a slight step in the middle of the parameter interval around $(M_L)_{ii}=2.5\times10^6~{\rm GeV}^2$ and
  remains stable in the rest of the parameter range.
  This indicates that $(M_L)_{ii}$ only provides subdominant corrections to $R_{\Lambda_c}$ and cannot dominate the deviation of the observable from the SM prediction.
  The numerical results of $\frac{R_{{\Lambda}_c}}{R^{SM}_{\Lambda_c}}$
  vary from 0.87 to 0.9.

  The Fig.\ref{fig6}(b) illustrates the behavior of the  observable $\frac{R_{{\Lambda}_c}}{R^{SM}_{\Lambda_c}}$ as a function of $(T_e)_{ii}$.
  Similar as Fig.\ref{fig4}(b), the curve exhibits step-like features.
   The both lines have a weak abrupt jump around
   the point $(T_e)_{ii}=0~{\rm GeV}$, when $(T_e)_{ii}$ varies  from -4000 GeV to 4000 GeV. They remain stable in the parameter regions $300 ~{\rm GeV}<|(T_e)_{ii}|<4000~{\rm GeV}$.   The jump direction of red dashed
   line($M_2$=6000 GeV) is opposite
   to that of the solid line ($M_2$=7000 GeV).
   This condition is similar as that of Fig.\ref{fig4}(b).

\subsection{Two-dimensional image}
To better study how parameters affect the observables, we also plot
 some multidimensional scatter plots, using the following parameters to create the images:
\begin{eqnarray}
(T_{\nu})_{ii}, ~(M_{E})_{ii},
~M_2, ~(Y_X)_{ii}, ~i=1,~2,~3.
\end{eqnarray}

 Similarly, we still analyze the presentation of the images.
\begin{figure}[ht]
\centering
\subfigure[]{
\setlength{\unitlength}{5.0mm}
\includegraphics[width=3.0in]{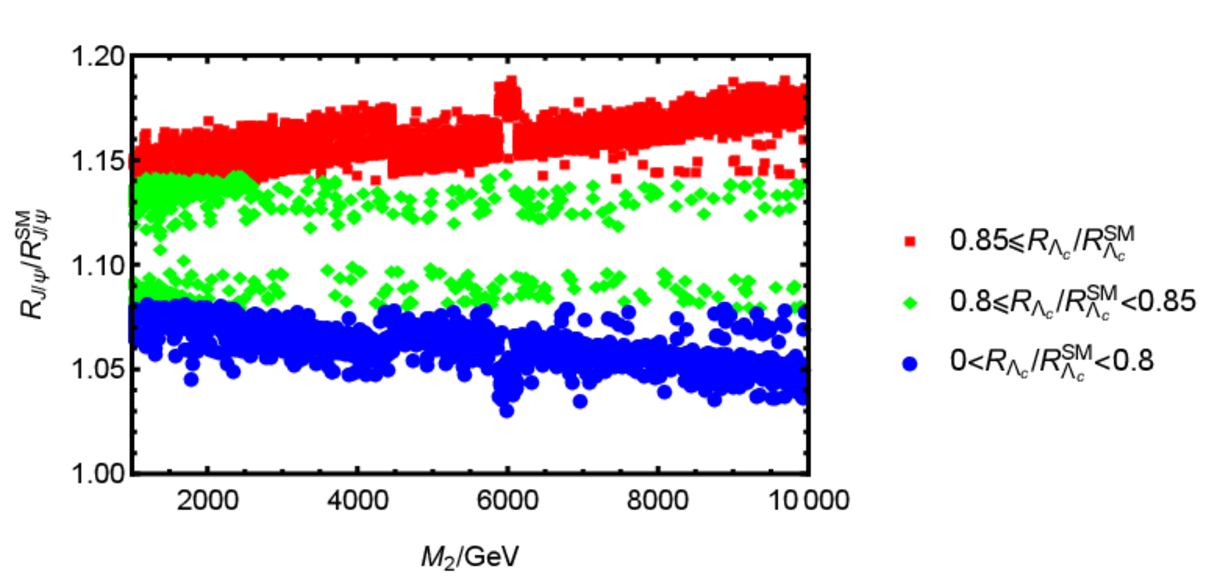}
\label{Fig20}
}
\subfigure[]{
\setlength{\unitlength}{5.0mm}
\includegraphics[width=3.0in]{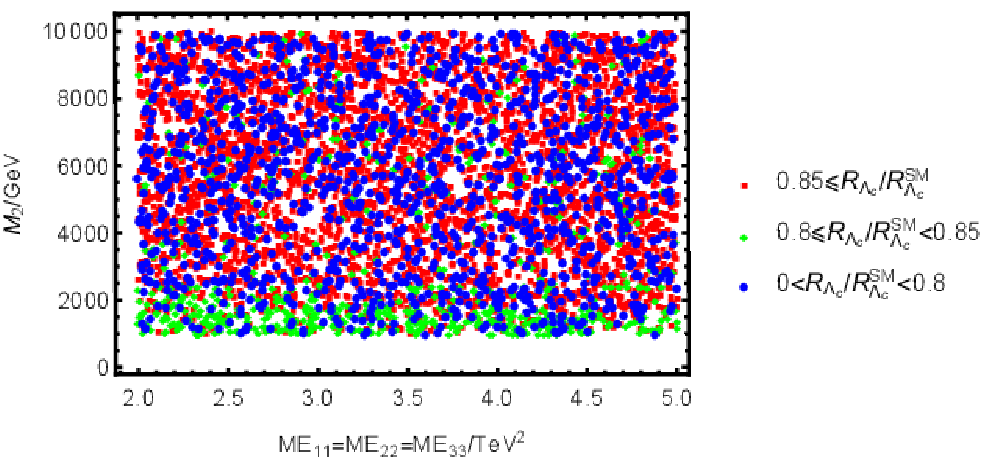}
\label{Fig21}
}
\caption{ The effects of $M_2$  and  $(M_{E})_{ii}$ on the observed quantity $\frac{R_{{\Lambda}_c}}{R^{SM}_{\Lambda_c}}$.} {\label {fig7}.}
\end{figure}

 To further explore the joint modulation mechanism of new physics parameters on  decay observables, we perform a scatter analysis for the $b\rightarrow c\tau\nu$ decay observables with the variables
 $(T_{\nu})_{ii}\in[-100,\,100]{\rm GeV}, ~(M_E)_{ii}\in[2.0,\,5.0]{\rm TeV}^2,
~M_2\in[1000,\,10000]{\rm GeV}, ~(Y_X)_{ii}\in[0.1,\,1], ~i=1,~2,~3$.
  Fig.\ref{fig7}(a) presents the distribution of parameter points on the $(M_2,\;R_{J/\psi}/R_{J/\psi}^{\rm SM})$ plane. Different colors are adopted to distinguish the value ranges of
  $\frac{R_{{\Lambda}_c}}{R^{SM}_{\Lambda_c}}$:

 ($\textcolor{red}{\blacksquare}$) represent $\frac{R_{{\Lambda}_c}}{R^{SM}_{\Lambda_c}}$ points within $\frac{R_{{\Lambda}_c}}{R^{SM}_{\Lambda_c}}$$\geq0.85$ range.

 ($\textcolor{green}{\blacklozenge}$) represent $\frac{R_{{\Lambda}_c}}{R^{SM}_{\Lambda_c}}$ points within $0.8\leq$$\frac{R_{{\Lambda}_c}}{R^{SM}_{\Lambda_c}}$$<0.85$ range.

 ($\textcolor{blue}{\bullet}$) represent $\frac{R_{{\Lambda}_c}}{R^{SM}_{\Lambda_c}}$ Points within $\frac{R_{{\Lambda}_c}}{R^{SM}_{\Lambda_c}}$$<0.8$ Range.

The numerical results indicate that the $\textcolor{red}{\blacksquare}$ points with a high ratio value are mainly concentrated in the top level. $\textcolor{green}{\blacklozenge}$ points
are in the middle and
$\textcolor{blue}{\bullet}$
 points are located in the the bottom layer.
For $M_2 \approx2000~{\rm GeV}$, the  values of
$R_{J/\psi}/R_{J/\psi}^{\rm SM}$ vary from 1.06 to 1.18.
When $M_2 \approx 10000~{\rm GeV}$, the  values of
$R_{J/\psi}/R_{J/\psi}^{\rm SM}$ extend from 1.04 to 1.20.
The shape of $\textcolor{red}{\blacksquare}$ $\textcolor{green}{\blacklozenge}$ $\textcolor{blue}{\bullet}$ implies that $M_2$
has visible effects to the numerical results.


Fig.\ref{fig7}(b) illustrates the sample distribution
 in the $(M_E)_{ii}$-$M_2$  plane.
 As discussed in the front, $M_2$ and $(M_E)_{ii}$
 affect the mass and mixings of slepton, chargino and neutralino.
 So we hope they produce some corrections.
 One can see that the $\textcolor{red}{\blacksquare}$, $\textcolor{green}{\blacklozenge}$ and $\textcolor{blue}{\bullet}$ samples are  mixed in the horizontal range $(M_E)_{ii}\in[2.0,5.0]{\rm TeV}^2$.
 $\textcolor{green}{\blacklozenge}$ points predominantly accumulate in the low-mass region $M_2\lesssim 2000{\rm GeV}$.
On the contrary, $\textcolor{blue}{\bullet}$ are spread throughout the whole area.
As $M_2$ increases, the decoupling effect from new-physics states gradually sets in, and the fraction of red samples rises
 significantly and dominates at large $M_2$.
 This behavior indicates that for $R_{\Lambda_c}/R_{\Lambda_c}^{\rm SM}$
the $SU(2)_L$ gaugino mass  $M_2$ is more sensitive than $(M_E)_{ii}$.

\begin{figure}[ht]
\centering
\subfigure[]{
\setlength{\unitlength}{5.0mm}
\includegraphics[width=3in]{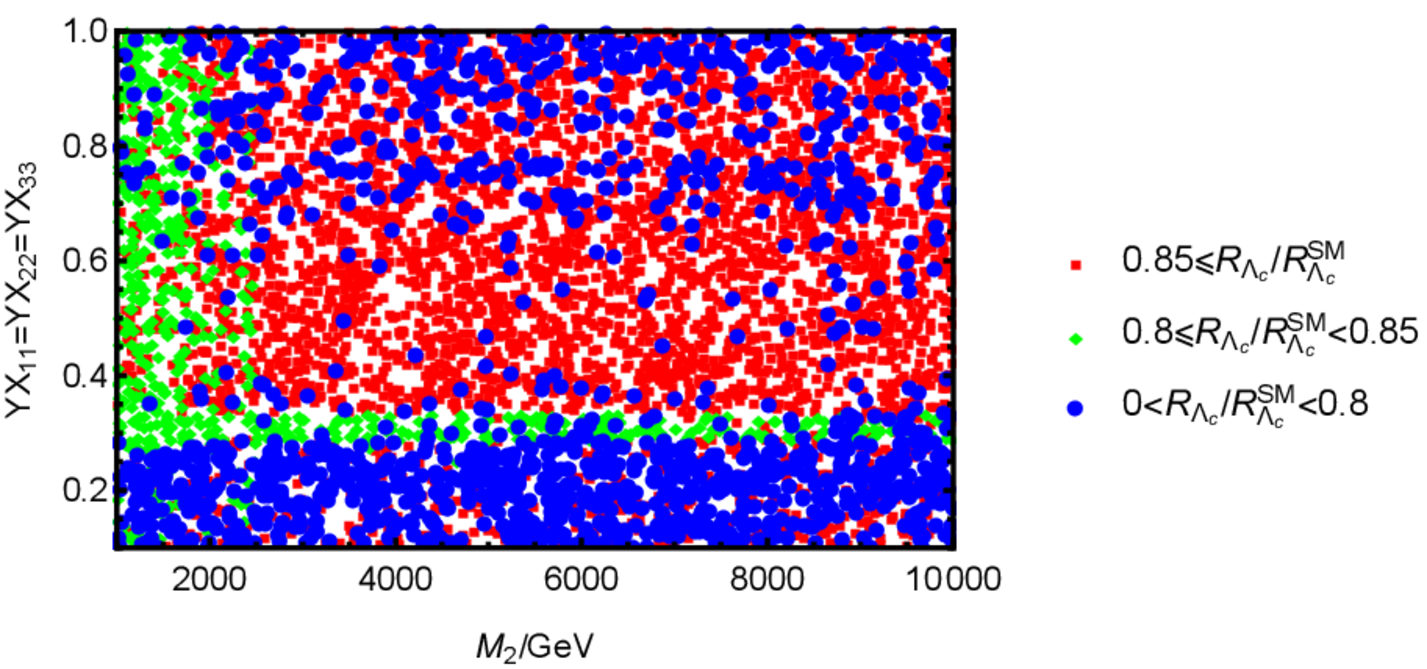}
\label{Fig22}
}
\subfigure[]{
\setlength{\unitlength}{5.0mm}
\includegraphics[width=3in]{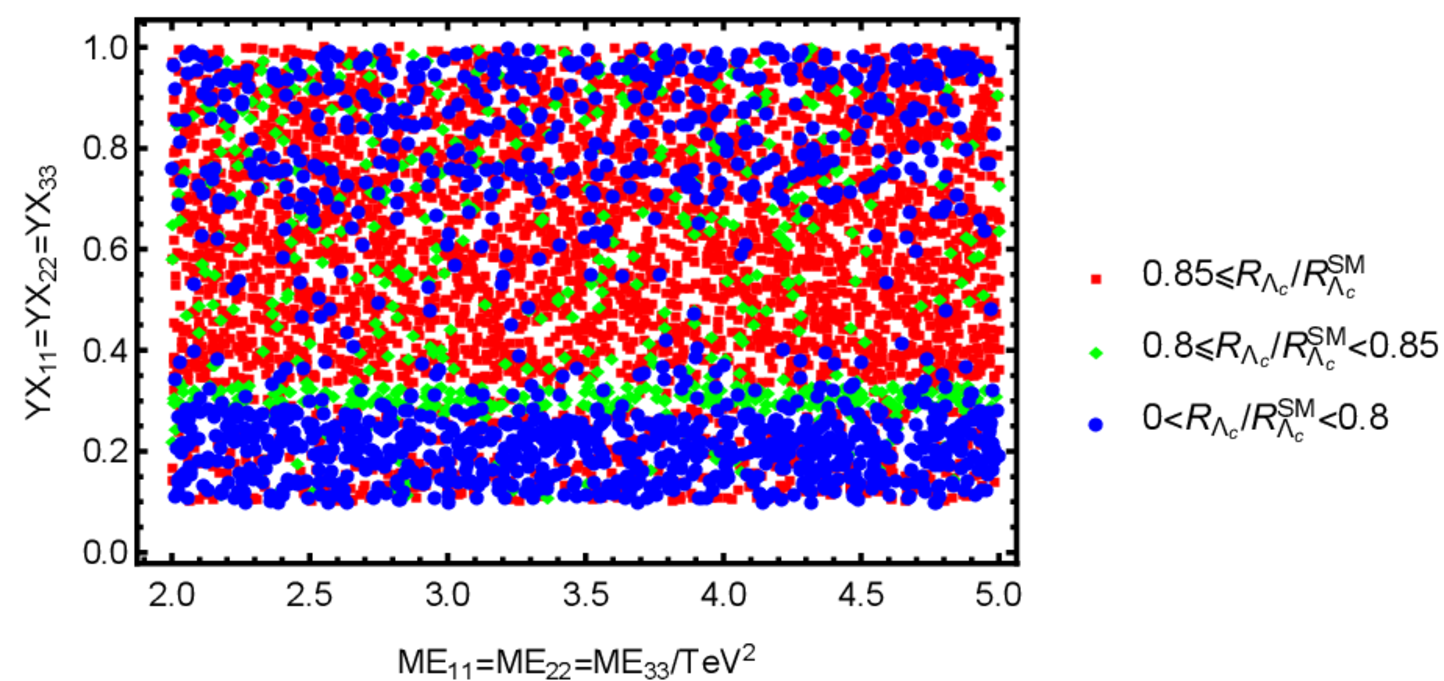}
\label{Fig23}
}
\subfigure[]{
\setlength{\unitlength}{5.0mm}
\includegraphics[width=3in]{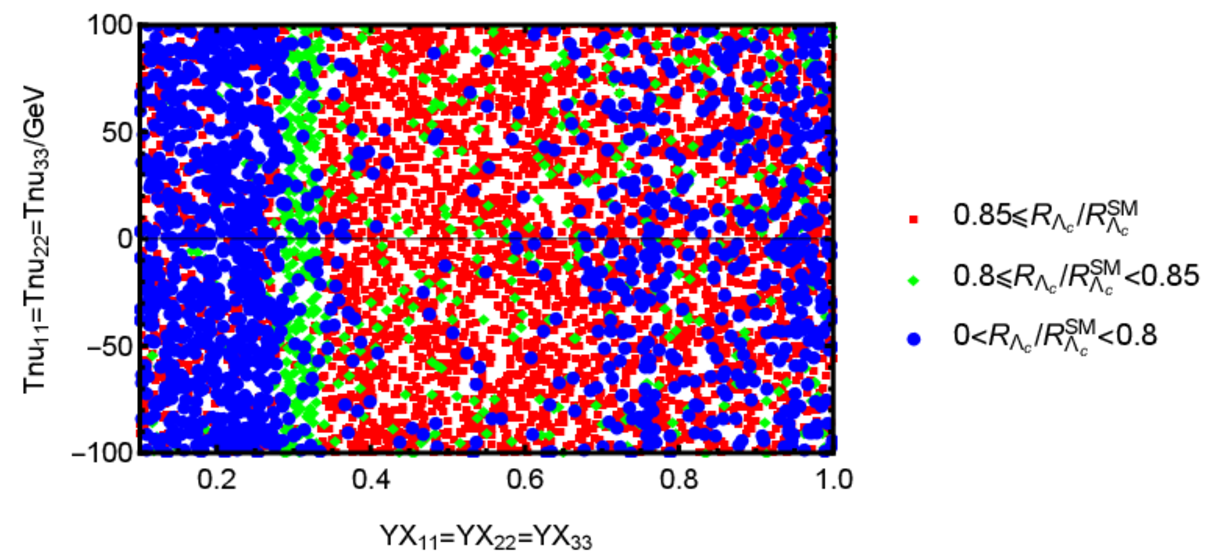}
\label{Fig24}
}
\caption{The effects of $M_2$, $(Y_X)_{ii}$, $(T_{\nu})_{ii}$ and  $(M_{E})_{ii}$  on the  quantity $\frac{R_{{\Lambda}_c}}{R^{SM}_{\Lambda_c}}$.} {\label {fig8}.}
\end{figure}

 The Fig.\ref{fig8}(a) shows the sample distribution
 on the $(M_2,(Y_X)_{ii})$ parameter plane.
 As a new parameter $(Y_X)_{ii}$
 appears in the neutrino and sneutrino mass matrix,
 which should affect the results to some extent.
 One observes a clear horizontal stratification of the three-color samples along the vertical axis. The $\textcolor{blue}{\bullet}$ points (strong suppression) mainly accumulate in the low-coupling region
$(Y_X)_{ii}\lesssim0.35$. $\textcolor{green}{\blacklozenge}$  points are mainly found in two areas.
 In one area, $\textcolor{green}{\blacklozenge}$ form a transition band around $(Y_X)_{ii}\approx0.3\sim0.4$. The other area is $M_2<2000{\rm GeV}$.
   While $\textcolor{red}{\blacksquare}$ points (weak suppression)
   occupy large upper region with
$(Y_X)_{ii}\gtrsim0.4$. This behavior indicates that $R_{\Lambda_c}/R_{\Lambda_c}^{\rm SM}$ is sensitive to the parameter $(Y_X)_{ii}$, whose controlling power is considerably stronger than that of
the gaugino mass $M_2$. Along the horizontal axis of $(M_E)_{ii}$ in the Fig.\ref{fig8}(b), the $\textcolor{red}{\blacksquare}$, $\textcolor{green}{\blacklozenge}$, and $\textcolor{blue}{\bullet}$ points are mixed almost
smoothly throughout the
 entire range of $2.0\sim5.0~{\rm TeV}^2$.
 It shows that $(M_E)_{ii}$ is dull parameter and affects the results weakly.
A pronounced stratification effect is
observed along the vertical axis of $(Y_X)_{ii}$. For $(Y_X)_{ii}$ near 0.3, the region is predominantly populated by $\textcolor{green}{\blacklozenge}$ points and form a narrow transition band. For $(Y_X)_{ii}\gtrsim0.35$,
$\textcolor{red}{\blacksquare}$ points emerge abundantly. In the high-$(Y_X)_{ii}$ region, $\textcolor{red}{\blacksquare}$ and $\textcolor{blue}{\bullet}$ samples coexist in an interlaced fashion.

   The Fig.\ref{fig8}(c) shows the sample distribution on the $(Y_X)_{ii}$-$(T_\nu)_{ii}$ parameter plane. $\textcolor{red}{\blacksquare}$, $\textcolor{green}{\blacklozenge}$, and $\textcolor{blue}{\bullet}$  denote the
   same  ranges as mentioned above. Along the horizontal axis $(Y_X)_{ii}$, the three sets of samples exhibit clear vertical stratification:   $\textcolor{green}{\blacklozenge}$ transition band appears for
   $(Y_X)_{ii}\approx 0.25\sim0.35$. Red samples appear in large numbers when $(Y_X)_{ii} \gtrsim 0.35$.  Blue samples appear in large numbers both when $(Y_X)_{ii} \lesssim 0.3$ and $(Y_X)_{ii} \gtrsim 0.7$, but they are a
   bit more densely distributed when $(Y_X)_{ii} \lesssim 0.3$. Overall, the range of
   $R_{\Lambda_c}/R_{\Lambda_c}^{\rm SM}$ is predominantly determined by $(Y_X)_{ii}$, and $(T_\nu)_{ii}$ only provides subdominant correction.

To explore the parameter space broader,
 we change a set of parameters and redraw the graphs again.
  The new parameters are as follows:
\begin{eqnarray}
(T_{e})_{ii}, ~(M_{L})_{ii},
~M_2, ~(M_{E})_{ii}, ~i=1,~2,~3.
\end{eqnarray}
To further explore the joint modulation mechanism of new physics parameters on  decay observables, we perform a scatter analysis for the $b\rightarrow c\tau\nu$ decay observables with the variables
$(M_L)_{ii}\in[2.0,\,5.0]{\rm TeV}^2, ~(M_E)_{ii}\in[2.0,\,5.0]{\rm TeV}^2,
~M_2\in[1000,\,10000]{\rm GeV}, ~(T_{e})_{ii}\in[-4000,\,4000]{\rm GeV}, ~i=1,~2,~3$.

\begin{figure}[ht]
\centering
\subfigure[]{
\setlength{\unitlength}{5.0mm}
\includegraphics[width=3in]{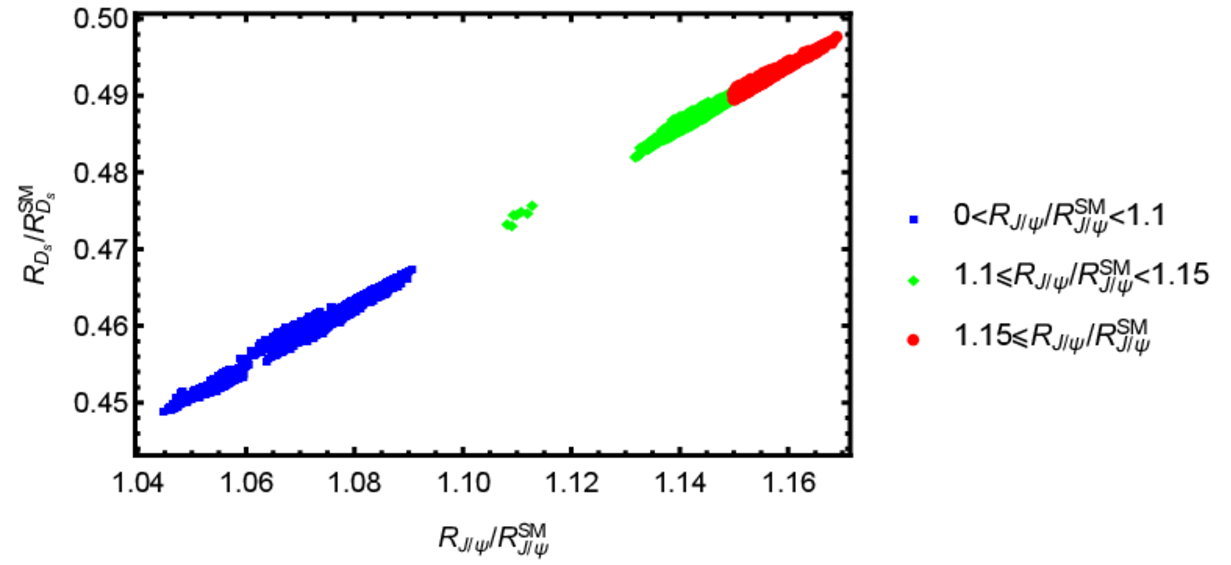}
\label{Fig25}
}
\subfigure[]{
\setlength{\unitlength}{5.0mm}
\includegraphics[width=3in]{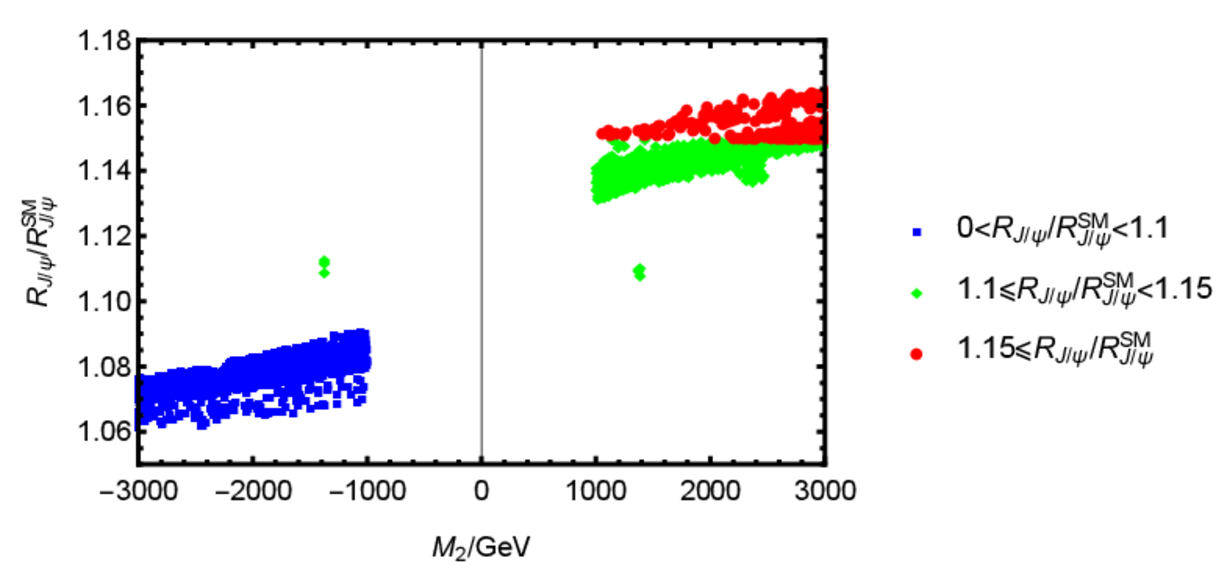}
\label{Fig26}
}
\subfigure[]{
\setlength{\unitlength}{5.0mm}
\includegraphics[width=3in]{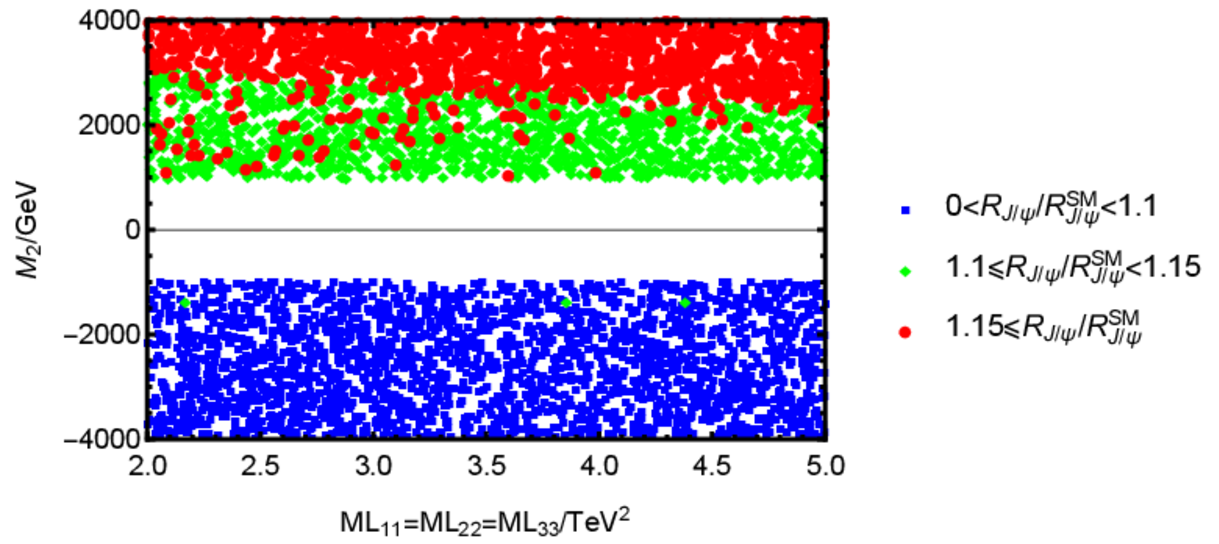}
\label{Fig27}
}
\subfigure[]{
\setlength{\unitlength}{5.0mm}
\includegraphics[width=3in]{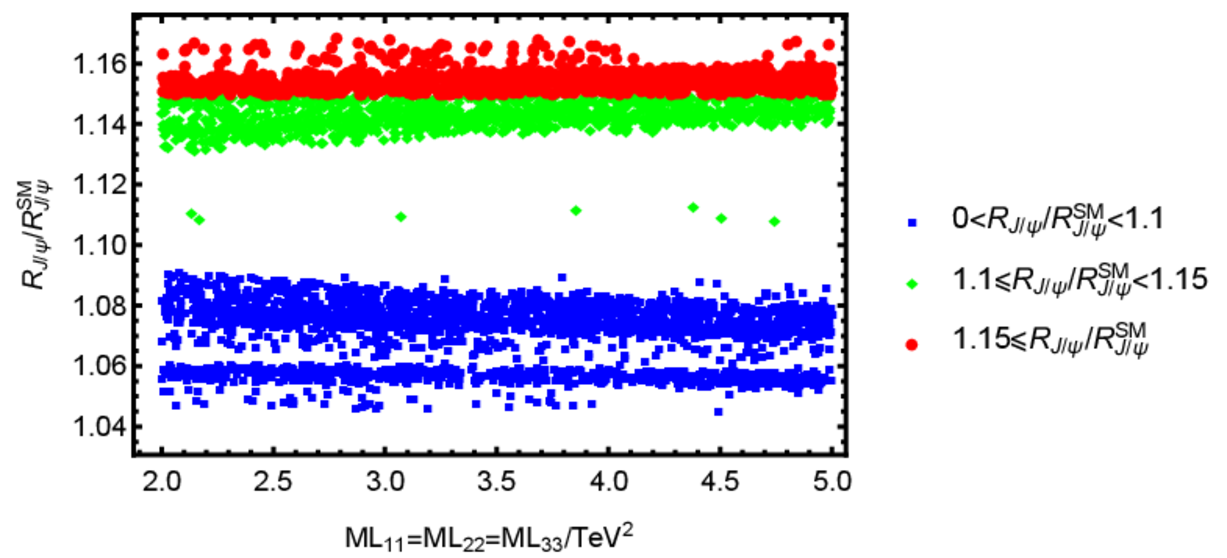}
\label{Fig28}
}
\caption{ With $M_2$  and  $(M_{L})_{ii}$  on the observed quantity $(\frac{R_{J/\psi}}{R^{SM}_{J/\psi}})$.} {\label {fig9}}
\end{figure}

 For the Fig.\ref{fig9}(a) in the \(R_{J/\psi}/R_{J/\psi}^{\rm SM}\) and \(R_{D_s}/R_{D_s}^{\rm SM}\) plane, we can observe that the numerical results cluster along a narrow inclined band. As \(R_{J/\psi}/R_{J/\psi}^{\rm
 SM}\) increases, \(R_{D_s}/R_{D_s}^{\rm SM}\) rises monotonically in parallel.

 ($\textcolor{red}{\bullet}$) represent $(\frac{R_{J/\psi}}{R^{SM}_{J/\psi}})$ points within $(\frac{R_{J/\psi}}{R^{SM}_{J/\psi}})$$\geq1.15$ range.

 ($\textcolor{green}{\blacklozenge}$) represent $(\frac{R_{J/\psi}}{R^{SM}_{J/\psi}})$ points within $1.1\leq$$(\frac{R_{J/\psi}}{R^{SM}_{J/\psi}})$$<1.15$ range.

 ($\textcolor{blue}{\blacksquare}$) represent $(\frac{R_{J/\psi}}{R^{SM}_{J/\psi}})$ points within $(\frac{R_{J/\psi}}{R^{SM}_{J/\psi}})$$<1.1$ range.

 In the Fig.\ref{fig9}(a), $\textcolor{blue}{\blacksquare}$ samples correspond to low values of both observables, $\textcolor{green}{\blacklozenge}$ samples lie in the intermediate range, and $\textcolor{red}{\bullet}$
 samples are associated
 with simultaneously large values for the two observables.
In the whole $(\frac{R_{J/\psi}}{R^{SM}_{J/\psi}})$ is near 1.
On the contrary,   \(R_{D_s}/R_{D_s}^{\rm SM}\) is around 0.5.
  This correlation indicates that, within the present model framework, the new-physics corrections to $B\rightarrow D_s\tau\nu$  cannot be fully decoupled, since both
 processes are subject to the constraints from the coupling structure of effective operators in different way.

  We present the sample distribution on the $(M_2,\,R_{J/\psi}/R_{J/\psi}^{\rm SM})$ plane in
 Fig.\ref{fig9}(b).
 The negative
 $M_2$ branch is dominated by $\textcolor{blue}{\blacksquare}$ samples, with $R_{J/\psi}/R_{J/\psi}^{\rm SM}$ concentrated in the range $1.06\sim1.09$.  In
 contrast, the positive $M_2$ branch is populated mainly by $\textcolor{green}{\blacklozenge}$ and $\textcolor{red}{\bullet}$ points, and the ratio rises to $1.10\sim1.17$ owing to constructive interference between
 new-physics and SM amplitudes. It leads to sizable enhancement. This behavior demonstrates that a sign flip of $M_2$ modifies the interference phase of effective couplings and qualitatively alters the direction of
 new-physics corrections to the observable. Within the positive branch, samples gradually shift from $\textcolor{green}{\blacklozenge}$ to $\textcolor{red}{\bullet}$ as $M_2$ increases. Overall, the sign of $M_2$ serves as
 the key factor governing the magnitude of new-physics corrections to $R_{J/\psi}$, while its absolute value only provides
 subdominant fine-tuning.

  The Fig.\ref{fig9}(c) displays the distribution of scanned samples on the $(M_L)_{ii}$ and $M_2$ parameter plane.
The positive-mass region $M_2>1000$ GeV accommodates abundant $\textcolor{green}{\blacklozenge}$, and $\textcolor{red}{\bullet}$ samples,
  allowing moderate to strong enhancement of $R_{J/\psi}$.
  By contrast, the negative mass region $M_2<-1000$ GeV is dominated by $\textcolor{blue}{\blacksquare}$ samples with only a tiny fraction of
  $\textcolor{green}{\blacklozenge}$ samples.
    Along the horizontal axis, for $(M_L)_{ii}$ within $2.0\sim5.0~{\rm TeV}^2$,
  samples of all colors spread uniformly without apparent stratification or selection effect.

   The Fig.\ref{fig9}(d) shows the sample distribution on the $(M_L)_{ii}$ and $R_{J/\psi}/R_{J/\psi}^{\rm SM}$ plane.  One observes three well-defined horizontal bands in the sample distribution: the
   $\textcolor{blue}{\blacksquare}$ band lies at $1.05\sim1.09$, the $\textcolor{green}{\blacklozenge}$ band at $1.13\sim1.15$, and the $\textcolor{red}{\blacksquare}$ band at
   $1.15\sim1.17$.
    Clear gaps separate different bands, revealing a discrete branch structure for the observable. Along the horizontal axis, the three bands extend uniformly over the full scanned range
   $(M_L)_{ii}\in[2.0,\,5.0]{\rm TeV}^2$, without visible tilting or positional shift.
   This behavior indicates that $(M_L)_{ii}$ barely modulates $R_{J/\psi}/R_{J/\psi}^{\rm SM}$ within the considered parameter space, and the
   discrete branches of the observables are predominantly determined by other new-physics parameters.

\begin{figure}[ht]
\centering
\subfigure[]{
\setlength{\unitlength}{5.0mm}
\includegraphics[width=3in]{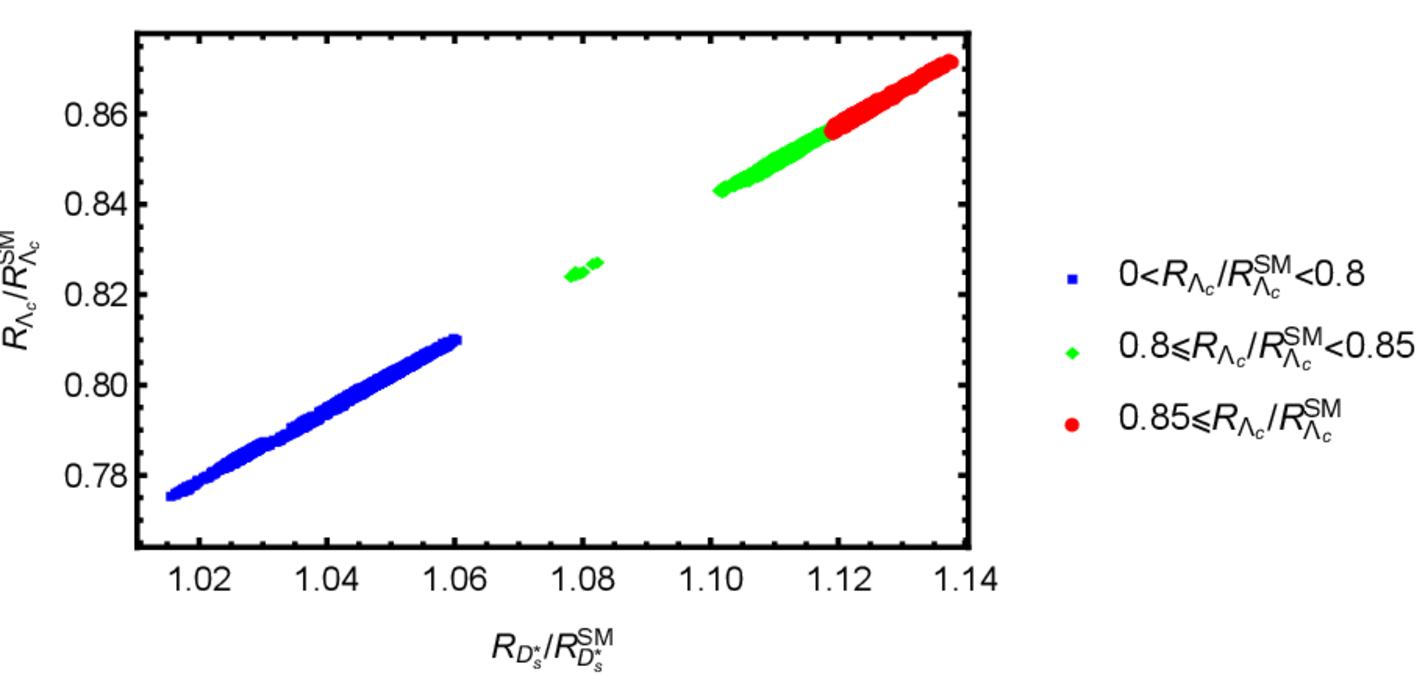}
\label{Fig29}
}
\subfigure[]{
\setlength{\unitlength}{5.0mm}
\includegraphics[width=3in]{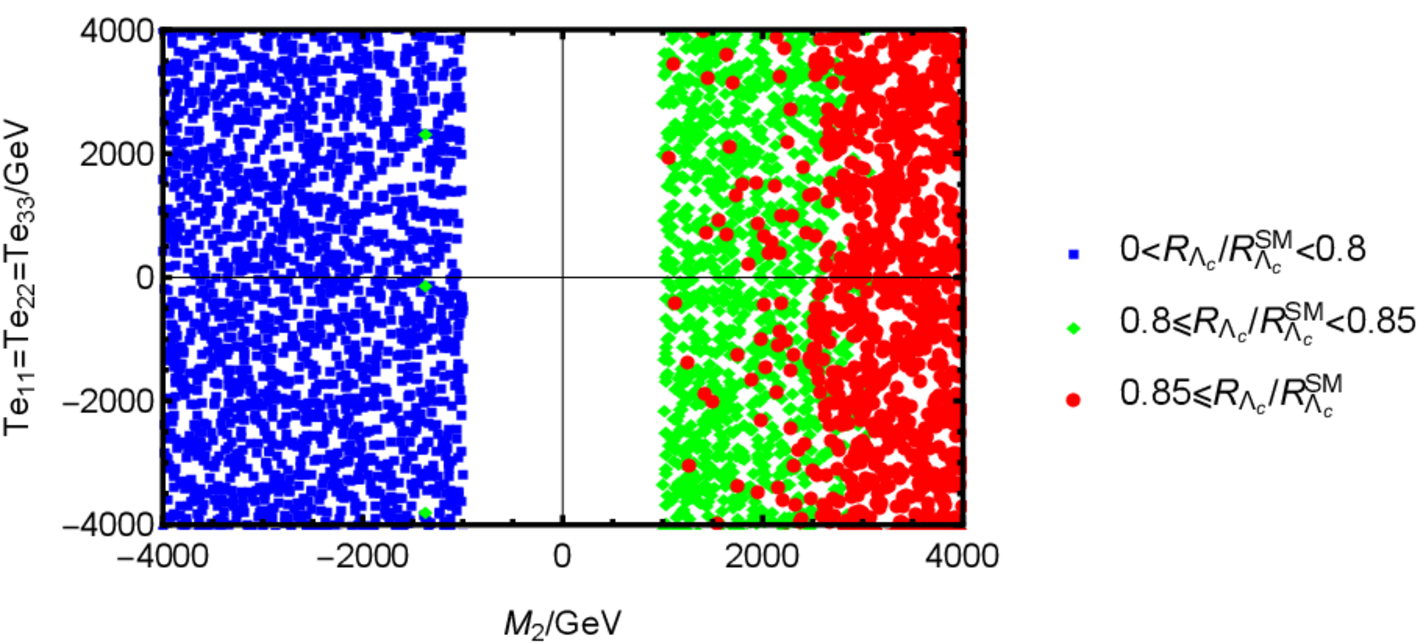}
\label{Fig30}
}
\caption{ The effects of $M_2$ and $(T_{e})_{ii}$  on the observed quantity $\frac{R_{{\Lambda}_c}}{R^{SM}_{\Lambda_c}}$.} {\label {fig10}}
\end{figure}

 In the Fig.\ref{fig10}, the notations are shown here

 ($\textcolor{red}{\bullet}$) represent $\frac{R_{{\Lambda}_c}}{R^{SM}_{\Lambda_c}}$ points within $\frac{R_{{\Lambda}_c}}{R^{SM}_{\Lambda_c}}$$\geq0.85$ range.

 ($\textcolor{green}{\blacklozenge}$) represent $\frac{R_{{\Lambda}_c}}{R^{SM}_{\Lambda_c}}$ points within $0.8\leq$$\frac{R_{{\Lambda}_c}}{R^{SM}_{\Lambda_c}}$$<0.85$ range.

 ($\textcolor{blue}{\blacksquare}$) represent $\frac{R_{{\Lambda}_c}}{R^{SM}_{\Lambda_c}}$ points within $\frac{R_{{\Lambda}_c}}{R^{SM}_{\Lambda_c}}$$<0.8$ range.

The samples exhibit a clear positive correlation behavior in Fig.\ref{fig10}(a):
 $R_{\Lambda_c}/R_{\Lambda_c}^{\rm SM}$ rises synchronously as $R_{D_s^*}/R_{D_s^*}^{\rm SM}$ increases. This correlation arises because both  processes are described by the same set of low-energy four-fermion
 effective operators and governed by identical new-physics Wilson coefficients.
  The parameter samples split into three inclined bands separated by empty gaps, manifesting the discrete-branch feature of the model.

 The Fig.\ref{fig10}(b) shows the shape of the numerical results spanned by  parameter $M_2$ and  $(T_e)_{ii}$. According to the image, we can see that the blue samples are concentrated around $M_2 < -1000 {\rm GeV}$, while
 the $\textcolor{red}{\bullet}$ and $\textcolor{green}{\blacklozenge}$ samples are distributed around $M_2>$ 1000 GeV. The stratification between the three samples is very obvious, but no stratification is seen in the
 vertical distribution, indicating that the parameter $(T_{e})_{ii}$ is a insensitive parameter.

\section{Discussion and Summary}
$U(1)_X$SSM is a $U(1)$ gauge extension of the MSSM and
has some advantages compared with WSSM.
We study the $b\rightarrow c l\nu$ process in the framework of $U(1)_X$SSM.
By performing systematic one-dimensional scans and two-dimensional sampling over the key new-physics parameters of the model, we comprehensively
 investigate how the new physics effects correct  the semileptonic $b\rightarrow c\tau\nu$ observables $R_{D_s}$, $R_{D_s^*}$, $R_{J/\psi}$ and $R_{\Lambda_c}$.

 All numerical
 results consistently demonstrate that the model parameters can be clearly classified into dominant control parameters and subdominant fine-tuning parameters.
Some sensitive parameters, like $l_W$ and $M_2$, can noticeably boost or suppress certain branches of the observable decays including the  $b\rightarrow c l\nu$ process.
For $R_{J/\psi}/R_{J/\psi}^{\rm SM}$ calculted under $U(1)_X$SSM, its value can reach around 1.3, getting closer to the experimental value,
 and there's a certain increase compared to the SM.
From the numerical analysis of $\frac{R_{{\Lambda}_c}}{R^{SM}_{\Lambda_c}}$,
 we find that its value can be lowered to around 0.85, which is somewhat suppressed compared to the SM. For the other two ratios without experimental data, the numerical results of $\frac{R_{{D^*}_s}}{R^{SM}_{D^*_s}}$ and
 $\frac{R_{{D}_s}}{R^{SM}_{D_s}}$ fluctuate near the SM predictions.

 Based on the numerical analysis in this article, we can see that, compared with the SM, the $U(1)_X$SSM with new gauge interactions and new particles can modify the Wilson coefficients of the $b\rightarrow c l\nu$ decay.
 It  offers a feasible theoretical explanation for the discrepancy between
experimental measurements and SM predictions.

\begin{acknowledgments}
This work is supported by National Natural Science Foundation of China (NNSFC)(No.12075074),
Natural Science Foundation of Hebei Province(A2020201002, A2023201040, A2022201022, A2022201017, A2023201041),
Natural Science Foundation of Hebei Education Department (QN2022173),
Post-graduate's Innovation Fund Project of Hebei University (HBU2024SS042),
This work is supported by the Project of the China Scholarship Council (CSC) No. 202408130113.
\end{acknowledgments}

\end{document}